\documentclass[12pt]{iopart}
\usepackage[hidelinks]{hyperref}
\usepackage{graphicx}
\graphicspath{{.}}
\usepackage{iopams}  
\begin{document}

\newcommand{\gsim}{\raisebox{-0.13cm}{~\shortstack{$>$ \\[-0.07cm]$\sim$}}~}

\title[Dynamics of a nonminimally coupled 
scalar field in asymptotically $AdS_4$ spacetime]{
Dynamics of a nonminimally coupled scalar 
field in asymptotically $AdS_4$ spacetime}

\author{Alex Pandya}
\address{Department of Physics, Princeton University, 
Princeton, New Jersey 08544, USA}
\ead{apandya@princeton.edu}
\vspace{10pt}
\author{Justin L. Ripley}
\address{DAMTP,
Centre for Mathematical Sciences,
University of Cambridge,
Wilberforce Road, Cambridge CB3 0WA, UK.
}
\ead{jr860@cam.ac.uk}
\vspace{10pt}

\begin{abstract}
   We numerically investigate the stability of four-dimensional
   asymptotically anti-de Sitter 
   ($AdS_4$) spacetime for a class of nonminimally coupled 
   scalar field theories.
   In particular, we study how the coupling affects 
   the formation of black holes, and the transfer of energy to different
   spatial/temporal scales. 
   We conclude by detailing the well-known analogy between the 
   nonminimally coupled scalar-field stress-energy tensor and that of a 
   viscous relativistic fluid, and discuss the limitations of that analogy 
   when it is applied to anisotropic scalar field configurations in asymptotically
   $AdS$ spacetimes.
\end{abstract}

\vspace{2pc}

\section{Introduction}
\label{sec:introduction}

Unlike asymptotically flat and asymptotically de Sitter spacetimes 
\cite{Friedrich1986,10.2307/j.ctt7zthns}, 
asymptotically anti de Sitter (AdS) spacetimes 
can be unstable to the formation of black holes, 
starting from arbitrarily weakly gravitating initial data.
Following a conjecture formulated by Dafermos and Holzegel 
\cite{Dafermos_unpublished},
Bizon and Rostworowski were the first to systematically (numerically)
study the nonlinear stability of asymptotically AdS spacetimes \cite{Bizon:2011gg}.
Their results strongly suggest that black holes form from generic, arbitrarily
small perturbations of the spacetime---at least for a massless
scalar field in spherical symmetry, in four spacetime dimensions 
($AdS_4$)\footnote{There are also classes of non-generic perturbations of AdS that
do not lead to black hole formation, constituting so-called ``islands of stability'' 
\cite{Buchel:2013uba,Balasubramanian:2014cja}.}.
Since then, there has been extensive work on numerically studying  
the AdS instability in other matter models and
in higher spacetime dimensions; see for example the reviews
\cite{Bizon:2013gxa,Evnin:2021buq}\footnote{Almost all numerical work on 
the instability of asymptotically AdS spacetimes has
been restricted to spherical symmetry, although a couple
recent studies have begun to relax these symmetry
assumptions---see \cite{Bantilan:2017kok,Bantilan:2020xas}.}.

In addition to numerical studies, perturbative calculations also suggest 
that asymptotically AdS spacetimes are unstable
to generic, small perturbations \cite{Dias:2011ss,Craps:2014vaa,Bizon:2015pfa}.
More recently, Moschidis has rigorously established the instability
of asymptotically $AdS_4$ spacetimes with a 
massless (null) Vlasov matter field 
\cite{Moschidis:2017llu,Moschidis:2018ruk}.
Heuristically, the instability of asymptotically AdS spacetimes is due
to the spacetime boundary being causally connected to the interior.  To have a 
well-posed initial value problem in these spacetimes, one must impose boundary 
conditions at spatial infinity, a natural choice for which
are Dirichlet (reflecting) boundary conditions 
\cite{Ishibashi:2004wx}
which prevent energy from escaping from the universe.  
As a result, instead of outgoing radiation dispersing to 
``infinity'',
it reflects back into the ``bulk'' of the spacetime, 
where it can gravitationally interact and eventually form a black hole. 

In this work we study the instability of asymptotically $AdS_4$ spacetimes
for a class of nonminimally coupled scalar field models.
Nonminimally coupled scalar fields are commonly found in low-energy
effective field theories approximating string theory 
\cite{Polchinski:1998rq,fujii_maeda_2003}, from which arose the
AdS/CFT correspondence \cite{Hubeny:2014bla}, 
which has motivated much of the current
interest in asymptotically AdS spacetimes.
It is unclear \textit{a priori} how a nonminimal coupling between 
the matter and gravitational sectors would impact the timescale to 
black hole formation,
thus studying these models may provide further insights into the 
nature of the instability of $AdS$ spacetime.

We derive the covariant equations of motion in
Sec.~\ref{sec:formulation}, and then describe the numerical
diagnostics we used in our code in Sec.~\ref{sec:diagnostics}.
In Sec.~\ref{sec:results} we describe our numerical results,
and discuss the well-known analogy between scalar fields and 
viscous fluids (as well as its limitations) in Sec.~\ref{sec:scalar_fluid_analogy}. 
Finally we summarize our results and suggest potential future work in Sec.~\ref{sec:discussion}.
In the Appendices we describe our numerical methodology in more detail,
provide convergence tests of our code, and present several longer
equations.

Throughout this study we use lowercase Latin letters to index spacetime 
tensor components, and use a ``mostly-plus'' signature $(-+++)$ for the 
spacetime metric. 

\section{\label{sec:formulation}Formulation of the equations of motion}
\subsection{Working in the Jordan frame}

In this work we consider the dynamics of a massless scalar field $\phi$ 
nonminimally coupled to Einstein gravity in four spacetime dimensions.  
This system is described by the action \cite{Kaiser2010,Winstanley2002}
\begin{equation} 
   \label{eq:Jordan_action}
   S 
   = 
   \int d^4x \, \sqrt{|g|} 
   \left[ 
      \frac{1}{2 \kappa} (R - 2 \Lambda) 
      - 
      \frac{1}{2} g^{ab} \nabla_a \phi \nabla_b \phi 
      - 
      \frac{\xi}{2} R \phi^2 
   \right],
\end{equation}
where $\kappa \equiv 8 \pi G/c^4$, $\Lambda$ is the 
cosmological constant (we use $\Lambda<0$ to set 
an asymptotically AdS spacetime geometry), 
and $\xi$ is the coupling constant between 
the scalar field and the Ricci scalar $R$.  
Note that setting $\xi = 0$ reduces the action to that 
of a minimally coupled scalar field.  
A scalar-tensor theory given by an action such as (\ref{eq:Jordan_action}) 
with an explicit coupling between $\phi$ and $R$ is said to be 
in the \textit{Jordan frame} \cite{Faraoni:1998qx,Flanagan:2004bz}.

Varying $S$ with respect to $\phi$ yields the scalar equation of motion,
\begin{equation} 
   \label{eq:Jordan_KGE}
   \nabla_a\nabla^a \phi = \xi R \phi
   .
\end{equation}
Varying with respect to the (inverse) metric $g^{a b}$ 
gives the Einstein equations \cite{Winstanley2002}
\begin{equation} 
   \label{eq:Jordan_EFE}
   G_{ab} = \kappa T_{ab}
   ,
\end{equation}
where the Einstein tensor 
$G_{ab} \equiv R_{ab} - \frac{1}{2} R g_{ab}$ and the stress-energy tensor 
is\footnote{Second covariant derivatives of $\phi$ appear in $T_{ab}$ due to 
	the nonminimal coupling term in the action; see, for example, Appendix C of 
	\cite{fujii_maeda_2003} for a detailed discussion.}
\begin{equation} 
   \label{eq:Jordan_Tab}
   \fl
   T_{ab} 
   \equiv 
   \frac{1}{1 - \kappa \xi \phi^2} \left[ 
      S_{ab} 
      - 
      2 \xi \left( 
         \phi \nabla_b \nabla_a \phi 
         + 
         \nabla_a \phi \nabla_b \phi 
         - 
         g_{ab} \left[
            \phi \nabla_c \nabla^c \phi 
            + 
            \nabla_c \phi \nabla^c \phi 
         \right] 
      \right) 
   \right]
   .
\end{equation}
Here $S_{ab}$ is the stress-energy tensor for a 
minimally coupled scalar field (which we define to include the cosmological constant term),
\begin{equation} 
   \label{eq:Jordan_Sab}
   S_{ab} 
   \equiv 
   \nabla_a \phi \nabla_b \phi 
   - 
   \frac{1}{2} g_{ab} \nabla_c \phi \nabla^c \phi 
   - 
   \frac{\Lambda}{\kappa} g_{ab}
   ,
\end{equation}
so $T_{ab} \to S_{ab}$ as $\xi \to 0$.

The presence of $R$ on the right-hand side of (\ref{eq:Jordan_KGE}) 
introduces terms proportional to second derivatives of the metric into the 
scalar evolution equation (which appear due to the total derivative that
appears in the variation of the Ricci scalar);
these may be removed in favor of terms proportional to first derivatives 
of $\phi$ using the trace of (\ref{eq:Jordan_EFE}) 
combined with (\ref{eq:Jordan_KGE}), which together imply \cite{Winstanley2002}
\begin{equation} 
   \label{eq:Jordan_R}
   R 
   = 
   \frac{
      4 \Lambda + \kappa (1 - 6 \xi) \nabla_c \phi \nabla^c \phi
   }{
      1 - \xi (1 - 6 \xi) \kappa \phi^2
   }.
\end{equation}
This result allows (\ref{eq:Jordan_KGE}) to be solved in 
much the same way as in the minimally-coupled case $\xi = 0$.

Despite this simplification,
the Einstein equations (\ref{eq:Jordan_EFE}-\ref{eq:Jordan_Sab}) 
are still complicated by the presence of second derivatives of $\phi$ 
and first derivatives of the metric
appearing in $T_{ab}$.  
Rather than engineer a numerical scheme capable of handling the 
Einstein equations in this form, 
we instead perform a Weyl transformation to solve the theory
in the \emph{Einstein frame}, where there is no direct coupling
between the scalar field and the Ricci scalar in the action.
\subsection{Transforming to the Einstein frame\label{sec:transform_einstein}}

To move to the Einstein frame, we perform a \emph{Weyl transformation}, 
i.e. a we apply a field redefinition of the form 
\begin{equation} 
   \label{eq:Weyl_rescaling}
   \hat{g}_{a b} \equiv \Omega^2 g_{ab}
   .
\end{equation}
Fields computed in the Einstein frame (in other words, computed using the
Weyl-rescaled metric $\hat{g}_{ab}$) will be written with hats.
Under the transformation (\ref{eq:Weyl_rescaling}) we have \cite{Kaiser2010}
\begin{eqnarray}
   \hat{g}^{ab} 
   &= 
   \Omega^{-2} g^{ab} 
   \nonumber\\
   \sqrt{|\hat{g}|} 
   &= 
   \Omega^4 \sqrt{|g|} 
   \label{eq:Einstein_R}
   \\
   \hat{R} 
   &= 
   \frac{1}{\Omega^2} \left[ R - \frac{6}{\Omega} \nabla^a\nabla_a \Omega \right]. 
   \nonumber
\end{eqnarray}
We then choose
\begin{equation} \label{eq:Omega_sq}
   \Omega^2 \equiv 1 - \kappa \xi \phi^2,
\end{equation}
so that the action (\ref{eq:Jordan_action}) becomes 
\begin{equation} \label{eq:intermediate_action}
   \fl
   S 
   = 
   \int d^4x \, 
   \left[ 
      \sqrt{|\hat{g}|} \frac{1}{2 \kappa} \hat{R} 
      + 
      \frac{3}{\kappa} g^{ab} \Omega \partial_a [\sqrt{|g|} \partial_b \Omega] 
      - 
      \sqrt{|g|} \frac{\Lambda}{\kappa} 
      - 
      \sqrt{|g|} \frac{1}{2} g^{ab} \nabla_a \phi \nabla_b \phi 
   \right].
\end{equation}
Integrating the second term by parts, 
noting that the boundary term vanishes\footnote{Note that
$\partial_b\Omega = -\kappa\xi\phi\partial_b\phi/\sqrt{1-\kappa\xi\phi^2}$.
As we impose Dirichlet ($\phi=0$) boundary conditions on the AdS boundary
(see Sec.~\ref{sec:coordinate_eom})
we conclude that the boundary term in the integration by parts of
the term in (\ref{eq:intermediate_action}) vanishes. 
See \cite{Ishibashi:2004wx} for more discussion.}, 
and rearranging further finally gives
\begin{equation} 
   \label{eq:Einstein_action}
   \fl
   S 
   = 
   \int d^4x \, \sqrt{|\hat{g}|} \left[ 
      \frac{1}{2 \kappa} \hat{R} 
      - 
      \left( 
         \frac{1 + (6 \xi - 1) \kappa \xi \phi^2}{(1 - \kappa \xi \phi^2)^2} 
      \right) 
      \frac{1}{2} \hat{g}^{ab} \hat{\nabla}_{a} \phi \hat{\nabla}_{b} \phi 
      - 
      \frac{\Lambda}{\kappa (1 - \kappa \xi \phi^2)^{2}} 
   \right],
\end{equation}
which does not have an explicit coupling term between 
$\phi$ and $\hat{R}$, unlike its Jordan-frame counterpart (\ref{eq:Jordan_action}).  
It is possible to redefine $\phi \to \hat{\phi}$ such that 
the scalar field's kinetic term in (\ref{eq:Einstein_action}) reduces 
precisely to minimally-coupled form, 
$\frac{1}{2} \hat{g}^{ab} \hat{\nabla}_{a} \hat{\phi} \hat{\nabla}_{b} \hat{\phi}$.  
This rescaling is less useful when there is a cosmological constant, as 
the redefinition $\phi \to \hat{\phi}$ significantly complicates the scalar
field coupling to the $\Lambda$ term\footnote{The presence of a nonzero 
	cosmological constant breaks the scale invariance of the action, 
	making it impossible to Weyl transform it into the action of a minimally 
	coupled, massless scalar field as can be done when $\Lambda = 0$.}. 
Because of this, we do not redefine the scalar field.

The scalar field equation of motion for the
Einstein frame action (\ref{eq:Einstein_action}) is
\begin{equation} \label{eq:Einstein_KGE}
   0 
   = 
   q(\phi) \hat{\nabla}_a \hat{\nabla}^a \phi 
   + 
   \frac{1}{2} \frac{d q(\phi)}{d \phi} 
   \hat{\nabla}_a \phi \hat{\nabla}^a \phi 
   - 
   \frac{dw(\phi)}{d\phi}
   .
\end{equation}
   The Einstein frame metric equation of motion is
\begin{equation} 
   \label{eq:Einstein_EFE}
   \hat{G}_{ab} 
   = 
   \kappa \hat{T}_{ab}
   .
\end{equation}
where the new stress-energy tensor is 
\begin{equation}
   \hat{T}_{ab} 
   \equiv
   q(\phi) \left[ 
      \hat{\nabla}_{a} \phi \hat{\nabla}_{b} \phi 
      - 
      \frac{1}{2} \hat{g}_{ab} \hat{\nabla}_{c} \phi \hat{\nabla}^{c} \phi 
   \right] 
   - 
   \frac{\Lambda \hat{g}_{ab}}{\kappa} 
   - 
   w(\phi) \hat{g}_{ab}
   .
\end{equation}
We have defined the shorthand
\begin{eqnarray}
   q(\phi) 
   &\equiv 
   \frac{1 + (6 \xi - 1) \kappa \xi \phi^2}{(1 - \kappa \xi \phi^2)^2} 
   ,
   \\
   w(\phi) 
   &\equiv 
   \left( 
      \frac{1}{(1 - \kappa \xi \phi^2)^2} 
      - 
      1 
   \right) 
   \frac{\Lambda}{\kappa} 
   .
\end{eqnarray}
Note that $\hat{T}_{ab}$ is free of metric derivative terms
and of second derivatives of the scalar field.

Ultimately what an observer ``sees'' will depend on how they couple to
the scalar field and to the spacetime metric.
While we solve the equations of motion in the Einstein frame,
we present measurements of, e.g. the spacetime mass, in the 
Jordan frame.
We emphasize that the Weyl transformation is a field redefinition
and is not change of coordinates: it leaves the field $\phi$ and 
derivatives $\partial_a$ unchanged.
\subsection{Coordinate equations of motion
\label{sec:coordinate_eom}
}

In the Einstein frame we adopt the same coordinate choice as in 
\cite{Bizon:2011gg}, $x^{a} = (t, x, \theta, \varphi)$, and use
the line element 
\begin{equation} 
   \label{eq:Einstein_metric_ansatz}
   d\hat{s}^2 
   = 
   \frac{\ell^2}{\cos^2 x} 
   \left( 
      -
      \hat{A} e^{-2 \hat{\delta}} dt^2 
      + 
      \frac{1}{\hat{A}} dx^2 
      + 
      \sin^2 x 
      \left[ 
         d\theta^2 
         + 
         \sin^2 \theta d\varphi^2 
      \right] 
   \right).
\end{equation}
In these coordinates, the time coordinate $t \in (-\infty, \infty)$, 
the compactified spatial coordinate $x \in [0, \frac{\pi}{2}]$, 
\begin{equation}
   \label{eq:def_ell}
   \ell^2 \equiv 
   -
   \frac{3}{\Lambda} 
   > 
   0
   ,
\end{equation} 
and a pure-$AdS_4$ spacetime is recovered when $\hat{A} = 1, \hat{\delta} = 0$.  
For the remainder of this work we restrict to spherical symmetry, 
so all quantities vary only in $(t, x)$.

Computing the Einstein frame equations of motion in the coordinate basis 
(\ref{eq:Einstein_metric_ansatz}) results in
\begin{eqnarray}
   \fl
   \hat{A}' 
   &= 
   - 
   \frac{1 + 2 \sin^2 x}{\sin x \cos x} (\hat{A} - 1) 
   - 
   \kappa \sin x \cos x \left( 
      \frac{1}{2} \hat{A}  
      \left[ \Phi^2 + \Pi^2 \right]q(\phi) 
      - 
      \frac{1}{\cos^2 x} 3 \xi z(\phi) 
   \right) 
   \label{eq:Einstein_Aprime_eqn}
   \\
   \fl
   \hat{\delta}' 
   &= 
   -
   \frac{1}{2} \kappa \sin x \cos x \, q(\phi) \left(\Phi^2 + \Pi^2 \right) 
   \label{eq:Einstein_delta_eqn}
   \\
   \fl
   \dot{\Pi} 
   &= 
   \frac{1}{\tan^2 x} \partial_x [e^{-\hat{\delta}} \tan^2 x \hat{A} \Phi] 
   + 
   \frac{1}{2} \frac{1}{q(\phi)} 
   \frac{d q(\phi)}{d \phi} 
   e^{-\hat{\delta}} \hat{A} \left( \Phi^2 - \Pi^2 \right) 
   + 
   \frac{e^{-\hat{\delta}}}{\cos^2 x} 
   \frac{3 \xi}{q(\phi)} 
   \frac{d z(\phi)}{d \phi} 
   \label{eq:Einstein_Pi_eqn}
   \\
   \fl
   \dot{\hat{A}} 
   &= 
   -
   \kappa \sin x \cos x \, \hat{A}^{2} e^{-\hat{\delta}} q(\phi) \Phi \Pi
   , 
   \label{eq:Einstein_Adot_eqn}
\end{eqnarray}
where we have introduced further shorthand for partial derivatives 
$\dot{f} \equiv \partial_t f, f' \equiv \partial_x f$, as well as
\begin{eqnarray}
   z(\phi) 
   &\equiv 
   \frac{w(\phi)}{\Lambda \xi}
   ,
   \\
   \Phi 
   &\equiv 
   \phi' \label{eq:Phi_defn} 
   ,
   \\
   \Pi 
   &\equiv 
   \dot{\phi} \sqrt{-\frac{g_{xx}}{g_{tt}}} 
   = 
   \dot{\phi} \sqrt{-\frac{\hat{g}_{xx}}{\hat{g}_{tt}}} 
   = 
   \hat{A}^{-1} e^{\hat{\delta}} \dot{\phi}
   . 
   \label{eq:Pi_defn}
\end{eqnarray}
Notice that both $\Phi, \Pi$ do not have hats since $\phi$, 
the coordinates $x^{a}$ defining the partial derivatives, 
and the combination $g_{xx}/g_{tt}$ 
are unchanged under the Weyl rescaling (\ref{eq:Weyl_rescaling}).
We note that taking $\xi \to 0$ implies $q(\phi) \to 1$ and $z(\phi), 
\frac{dq}{d\phi}, \frac{dz}{d\phi} \to 0$.
In this limit, 
the system (\ref{eq:Einstein_Aprime_eqn}-\ref{eq:Einstein_Adot_eqn}) 
reduces to that solved in \cite{Bizon:2011gg}.

Taylor series expanding $\phi, \hat{A}, \hat{\delta}$ about $x = 0$, 
substituting into the equations of motion 
(\ref{eq:Einstein_Aprime_eqn}-\ref{eq:Einstein_Adot_eqn}),
and imposing regularity of the scalar field and the metric 
(i.e. that they are nonsingular, at least at early times) 
implies that near the origin
\begin{eqnarray} \label{eq:origin_falloff}
   \phi(t,x) 
   = 
   \phi(t,0) + \mathcal{O}(x^2)
   , 
   \nonumber\\
   \hat{\delta}(t,x) 
   = 
   \delta(t,0) + \mathcal{O}(x^2)
   , 
   \\
   \hat{A}(t,x) 
   = 
   1 + \mathcal{O}(x^2)
   \nonumber
   ,
\end{eqnarray}
which implies the boundary conditions 
$\Phi = \Pi' = \hat{A}' = \hat{\delta}' = 0$, and $A = 1$ at $x = 0$.  
We next consider the AdS boundary. We define $\rho \equiv \frac{\pi}{2} - x$,
expand the equations of motion about $\rho\ll1$, and impose 
that (\ref{eq:Einstein_Aprime_eqn}-\ref{eq:Einstein_Adot_eqn}) are regular
to obtain
\begin{eqnarray}
   \label{eq:bc_fields}
   \phi(t, x) 
   = 
   \phi_{\infty}\left(t\right)\rho^{\Delta}
   +
   \mathcal{O}\left(\rho^5\right)
   , 
   \nonumber\\
   \hat{\delta}(t,x) 
   = 
   \delta_{\infty}\left(t\right) 
   + 
   \mathcal{O}\left( \rho^{2\Delta} \right)
   , 
   \\
   \hat{A}(t,x) 
   = 
   1 
   -
   2M\rho^3
   +
   \mathcal{O}\left( \rho^{2\Delta-1}\right)
   \nonumber,
\end{eqnarray}
where 
\begin{equation}
   \Delta
   \equiv
   \frac{3}{2}
   +\left[
      \left(\frac{3}{2}\right)^2
      +
      \ell^2m_{eff}^2
   \right]^{1/2}
   \equiv
   \frac{3}{2}
   +\left[
      \left(\frac{3}{2}\right)^2
      -
      12\xi
   \right]^{1/2}
   .
\end{equation}
Here we have defined the scalar field effective mass
(see Sec.~\ref{sec:jordan_frame_lin_mode_energy} for more discussion)
\begin{equation}
   \label{eq:meff_definition}
   m_{eff}^2
   \equiv
   -
   12\frac{\xi}{\ell^2}
   =
   4\xi\Lambda
   .
\end{equation}
We have also made use of the constant $M$ in (\ref{eq:bc_fields}), which is the 
Einstein frame Misner-Sharp(-Hawking-Hayward) mass function evaluated at 
$\rho=0$; see Eq.~(\ref{eq:einstein_frame_mass}).
The conditions (\ref{eq:bc_fields}) imply that at the AdS boundary
$x = \frac{\pi}{2}$, one has
$\Phi = \Pi = \hat{A}' = \hat{\delta}' = 0$, 
and $\hat{A} = 1$.  

The metric ansatz (\ref{eq:Einstein_metric_ansatz}) possesses an 
additional degree of gauge freedom in that a shift 
$\hat{\delta} \to \hat{\delta} + C$ can be absorbed by suitable 
redefinition of $t$; we fix this gauge freedom by setting 
$\hat{\delta}(t, \frac{\pi}{2}) = 0$ as in \cite{Buchel:2012uh}, 
so $t$ reduces to the proper time for an observer at the boundary.
We see that we can have regular solutions at the AdS boundary so long as 
\begin{equation} 
   \label{eq:BF_bound}
   \ell^2m_{eff}^2 \geq -\frac{9}{4} 
   \iff 
   \xi \leq \frac{3}{16}
   ,
\end{equation}
which in the context of massive scalar fields in AdS spacetime 
is known as the \textit{Breitenlohner-Freedman} (BF) bound 
\cite{BREITENLOHNER1982197,BREITENLOHNER1982249}.  

All of the unknowns ($\hat{A}, \hat{\delta}, \phi$) 
are completely constrained by solving the three equations 
(\ref{eq:Einstein_Aprime_eqn}-\ref{eq:Einstein_Pi_eqn}); 
the remaining equation, (\ref{eq:Einstein_Adot_eqn}), 
may be used to as an additional check on the accuracy of the solution.
For the sake of brevity, we leave a detailed description of the numerical 
method used to solve (\ref{eq:Einstein_Aprime_eqn}-\ref{eq:Einstein_Pi_eqn}) 
to \ref{sec:numerical_methods}.

\section{Diagnostics\label{sec:diagnostics}}
We are interested in studying the dynamics of the Einstein-Klein Gordon system 
(\ref{eq:Einstein_Aprime_eqn}-\ref{eq:Einstein_Adot_eqn}) in light of the 
weakly turbulent instability of asymptotically AdS spacetime 
\cite{Bizon:2011gg}.  
In particular, our main goal is to investigate the dependence of the 
instability on the coupling constant $\xi$.  
For the sake of clarity, we make all comparisons in the Jordan frame, 
where $\xi$ serves the role of a coupling constant between curvature 
and the scalar field.
In the Einstein frame, $\xi$ plays a more 
complicated role, appearing in the scale factor 
$\Omega^2$ (\ref{eq:Weyl_rescaling}-\ref{eq:Omega_sq}) defining 
the Weyl transformation;
hence different values of $\xi$ define different 
Einstein frames, and 
comparisons across frames would be significantly more
difficult.

\subsection{Secular growth of gradients}

The AdS instability arises due to the transfer of energy from 
large spatial/temporal scales to small ones, resulting in an 
increase in spacetime curvature eventually cut off by the formation
of a black hole horizon.  
To monitor the growth of gradients, we compute three quantities at the 
origin of our spherically symmetric coordinate system: $\Pi^2$, 
which is equivalent between the Jordan and Einstein frames; 
the Jordan-frame Ricci scalar; and the Jordan-frame Kretschmann scalar 
$K \equiv R_{a b c d} R^{a b c d}$.  
Computing $\Pi^2$ is trivial, as we dynamically evolve $\Pi$ in 
the numerical algorithm.

To compute the Jordan-frame curvature scalars, one may begin by  
noting that our metric ansatz in the Einstein frame 
(\ref{eq:Einstein_metric_ansatz}) combined with the Weyl rescaling 
(\ref{eq:Weyl_rescaling}) defines the Jordan-frame metric tensor:
\begin{equation} 
   \label{eq:Jordan_metric}
   g_{ab} 
   = 
   \Omega^{-2} \hat{g}_{ab} 
   = 
   \Omega^{-2} \frac{\ell^2}{\cos^2 x} 
   \textnormal{diag}\left( 
      -\hat{A} e^{-2 \hat{\delta}}, \,
      \frac{1}{\hat{A}}, \,
      \sin^2 x, \,
      \sin^2 x \sin^2 \theta 
   \right)
   .
\end{equation}
As mentioned previously, the coordinates themselves 
${x^{a} = (t, x, \theta, \varphi)}$ are unchanged between the 
Einstein and Jordan frames, so coordinate derivatives are unchanged as well.
From (\ref{eq:Jordan_R}) 
one can now compute the Jordan-frame Ricci scalar
\begin{equation} \label{eq:Jordan_R_in_basis}
   R 
   = 
   \frac{
      -12 
      + 
      \kappa (1 - 6 \xi) \Omega^{2} \hat{A} \cos^2 x \, [\Phi^2 - \Pi^2]
   }{
      \ell^2 [1 - \xi (1 - 6 \xi) \kappa \phi^2]
   }. 
\end{equation}
Note that at the origin $\xi \to 0$ implies 
$R(x=0) \to \frac{-12 - \kappa \Pi^2}{\ell^2}$, 
in agreement with \cite{Bizon:2011gg} 
(after noting that they take $\kappa = 2$).

From inspection of (\ref{eq:Jordan_R}) or (\ref{eq:Jordan_R_in_basis}) it is 
clear that the Jordan-frame Ricci scalar is constant for the case of 
conformal coupling, $\xi = \frac{1}{6}$, and thus cannot provide any 
information about the transfer of energy in that case.  
Hence a different curvature scalar is required to make comparisons across
all $\xi$, so we also compute the Jordan-frame Kretschmann scalar 
$K \equiv R_{a b c d} R^{a b c d}$.  
The expression for $K$ even when evaluated at the origin $x =  0$ 
is extremely lengthy, so we display it in \ref{sec:Kretschmann}.

\subsection{Jordan-frame linearized mode energy
\label{sec:jordan_frame_lin_mode_energy}
}

Another diagnostic is to evaluate the spatial concentration of 
energy as a function of time. We measure this by 
projecting the solution for the scalar field (or its derivatives) onto 
an orthonormal basis of functions with a suitable notion of energy per 
basis function.  Such a basis can be found in the limit $\phi \to 0$ 
(or equivalently the limit where the geometry decouples from 
the matter content, $\kappa \to 0$).  
In this case, one has a scalar field propagating on a fixed
$AdS_4$ spacetime, and the wave equation (\ref{eq:Jordan_KGE}) becomes
\begin{equation} \label{eq:decoupled_KGE}
   \nabla_a\nabla^a \phi 
   = 
   -
   \frac{12}{\ell^2} \xi \phi
   ,
\end{equation}
since the pure-$AdS_4$ solution has $R = -\frac{12}{\ell^2}$.  
Note that (\ref{eq:decoupled_KGE}) takes the form of a 
massive scalar wave equation with mass $m_{eff}$; see
(\ref{eq:meff_definition}).

The decoupled scalar wave equation (\ref{eq:decoupled_KGE}) 
has a complete basis of orthonormal solutions given by 
\cite{PhysRevD.18.3565,Fodor:2013lza,Fodor:2015eia}
\begin{equation} \label{eq:mode_solutions}
   e_{j}(t,x) 
   = 
   C_{j} \cos \left( [\Delta + 2 j] t \right) 
   (\cos x)^{\Delta} 
   P_{j}^{1/2, \Delta-3/2} \left[ \cos(2 x) \right],
\end{equation}
which are labeled by the (non-negative integer) index $j$.  
The quantity $C_{j}$ is the scalar field amplitude which is 
fixed by requiring the solutions to be orthonormal with respect 
to the norm given in (\ref{eq:inner_product}), 
and $P^{\alpha,\beta}_{j}$ are Jacobi polynomials
(our conventions follow \cite{NIST:DLMF}).

Using the orthogonal basis of mode solutions 
(\ref{eq:mode_solutions}) one can define a measure of the mode energy 
which gives the support of a given mode $e_{j}$ in 
the solution at a given time.  
One such definition is
\begin{equation} 
\label{eq:mode_projections}
\Phi_{j} 
\equiv 
\Bigg\langle  
\Phi \sqrt{\frac{\ell^2 \Omega^{-2}}{\cos^2 x} g^{xx}} 
\, \Bigg| \, 
e_{j}' 
\Bigg\rangle
, 
~~~ 
\Pi_{j} 
\equiv 
\Bigg\langle 
\Pi \sqrt{\frac{\ell^2 \Omega^{-2}}{\cos^2 x} g^{xx}} 
\, \Bigg| \, 
e_{j} 
\Bigg\rangle
,
\end{equation}
where the prime denotes a spatial derivative $\partial_x$, 
and the inner product is defined as
\begin{equation} 
   \label{eq:inner_product}
   \langle f(x) | g(x) \rangle 
   \equiv 
   \int_{0}^{\frac{\pi}{2}} f(x) g(x) \tan^2 x \, dx,
\end{equation}
which is defined so that the mode solutions (\ref{eq:mode_solutions}) 
are orthonormal, 
$\langle e_{j} | e_{k} \rangle = \delta_{j, k}$.
The projections (\ref{eq:mode_projections}) are defined with the 
prefactor
$\sqrt{\frac{\ell^2 \Omega^{-2}}{\cos^2 x} g^{xx}}
=
\sqrt{\frac{\ell^2 }{\cos^2 x} \hat{g}^{xx}}$ so that they 
reduce in the minimal coupling limit $\xi \to 0$ to the definitions 
provided in \cite{Bizon:2011gg,Maliborski:2013via}. 
We define the energy per mode to be
\begin{equation} \label{eq:mode_energy}
   E_{j} 
   =
   \Pi_{j}^{2} 
   + 
   \omega_{j}^{-2} \Phi_{j}^2
   ,
\end{equation}
where $\omega_{j} \equiv \Delta + 2 j$.  
The mode energies $E_j$ are related to a measure of the total energy by
\begin{equation}
   E 
   = 
   \frac{1}{2} \int_{0}^{\frac{\pi}{2}} \Big( 
      \frac{\ell^2 \Omega^{-2}}{\cos^2 x} g^{xx} 
   \Big) 
   (\Pi^2 + \Phi^2) \tan^2 x 
   \, 
   dx 
   = 
   \sum_{j=0}^{\infty} E_{j}
\end{equation}
again defined entirely in the Jordan frame, 
such that it agrees with \cite{Bizon:2011gg,Maliborski:2013via} for $\xi = 0$.

It is important to note that projecting the solution onto the mode basis 
(\ref{eq:mode_solutions}) works best when using a set of initial data 
with compact support, so that there are times when the field vanishes at 
the AdS boundary.  This is because the basis functions $e_{j}$ 
have a different asymptotic fall-off as $x \to \frac{\pi}{2}$ 
than does the self-gravitating scalar field \cite{Maliborski:2013via}.  
To avoid such problems, we only use $E_j$ as a diagnostic for compactly 
supported initial data at times when the field is imploding through the 
origin (and thus vanishingly small near the AdS boundary).

\subsection{Apparent horizon radius and Misner-Sharp mass
\label{sec:ah_and_ms}}

To investigate the strong-field dynamics of the nonminimally coupled 
scalar field in asymptotically $AdS_4$ spacetime, we compute the location
of trapped surface formation in the Jordan frame.
We expect that after a trapped surface forms, the spacetime eventually
asymptotes to a Schwarzschild-$AdS_{4}$ spacetime \cite{Schleich2010}
(note that our choice of coordinates does not allow us to evolve exactly
to or past apparent horizon formation, either in the Jordan or
Einstein frame; see below).
One potential complication to this claim is that
in the Jordan frame the Null Convergence Condition 
(NCC; $R_{ab}k^ak^b\geq0$ for all null $k^a$)--which is one of the conditions
used in the proof of the classical black hole area law 
\cite{hawking_area_original}--can be violated.
We see this by contracting the
tensor equations of motion in that frame with a null vector $k^a$: 
\begin{equation}
R_{ab} k^a k^b = \frac{\kappa}{1 - \kappa \xi \phi^2}\left(\left(1-2\xi\right)\left(k^a\nabla_a\phi\right)^2-2\xi\phi k^ak^b\nabla_a\nabla_b\phi\right).
\end{equation}
If $k^ak^b\nabla_a\nabla_b\phi$ is sufficiently large
(negative or positive, depending on the sign of $\xi$), it is possible
for the NCC to be violated in the Jordan frame.
By contrast, the NCC always holds in the Einstein frame for this class
of theories
(at least so long as there is a nonsingular Weyl 
transformation between frames).
To show this, we contract a null vector $\hat{k}^a$ 
with the Einstein frame equations of motion to obtain
\begin{equation}
\hat{R}_{ab} \hat{k}^a \hat{k}^b = \kappa \frac{1+(6\xi-1)\kappa\xi\phi^2}{\left(1-\kappa\xi\phi^2\right)^2} \left(\hat{k}^a\hat{\nabla}_a\phi\right)^2.
\end{equation}
If $\xi<0$, this is clearly positive definite. If $\xi>0$, 
having $1+(6\xi-1)\kappa\xi\phi^2<0$ implies
$1-\kappa\xi\phi^2<0$, which would mean the denominator
$(1-\kappa\xi\phi^2)^2(=\Omega^2)$ would have to have gone through 
zero, which would make the Weyl transformation singular\footnote{As $\phi=0$
at the AdS boundary, we cannot have $\kappa\xi\phi^2>1$ throughout 
the entire spacetime.}.
We do not expect for potential violations of the NCC in the Jordan frame
to be able to reverse the formation
of a black hole once it has formed\footnote{This expectation holds true
at least for Brans-Dicke
gravity in asymptotically flat spacetimes \cite{Scheel:1994yr,Scheel:1994yn}.},
although it is beyond the scope of this work to verify this assumption.
With this potential caveat in mind, we next describe how we compute the location
of apparent horizons in the Jordan frame.

In spherical symmetry, we can write the outward null expansion as 
\cite{baumgarte_shapiro_2010}
\begin{equation}
   \label{eq:apparent_horizon_general_k}
   \Theta_{(k)}
   \equiv
   \frac{1}{4\pi R^2_{area}}
   k^{a}\nabla_{a}
   \left(4\pi R_{area}^2\right)
   ,
\end{equation}
where $R_{area}$ is the areal radius and $k^{a}$ is an outward pointing
null vector. From (\ref{eq:Jordan_metric}) we see that the areal radius in
the Jordan frame is
\begin{equation}
   R_{area}
   =
   \frac{1}{\Omega} \ell \tan x
   .
\end{equation}
Substituting the outward-pointing null vector
\begin{equation}
   k^a
   =
   \left(1,\hat{A}e^{-\hat{\delta}},0,0\right)
   .
\end{equation}
into Eq.~(\ref{eq:apparent_horizon_general_k}),
we see that an apparent horizon in the Jordan frame is 
then located at the points where the following relation is satisfied:
\begin{equation}
   \Theta_{(k)}
   =
   2\hat{A}e^{-\hat{\delta}}\left(
      \frac{1}{\sin x\cos x}
      +
      \frac{\kappa\xi\phi\left(\Phi+\Pi\right)}{1-\kappa\xi\phi^2}
   \right)
   =
   0
   .
\end{equation}

We compute the Jordan frame Misner-Sharp mass \cite{Misner:1964je}
(also called the Hawking-Hayward mass \cite{Hawking:1968qt,Hayward:1993ph})
through the following definition \cite{Maeda:2007uu,Fodor:2015eia}
\begin{equation}
1 - \frac{2M_{MS}}{R_{area}} - \frac{1}{3}R_{area}^2\Lambda \equiv g^{ab} \nabla_{a} R_{area} \nabla_{b} R_{area}
.
\end{equation}
Substituting in for $R_{area}$ and $\Lambda$, we see that 
\begin{eqnarray}
\label{eq:misner_sharp_mass}
   \fl
   M_{MS}
   =
   \frac{\ell}{2}\frac{\tan x}{\sqrt{1-\kappa\xi\phi^2}}\Bigg(
      &\frac{\sec^2x - \kappa\xi\phi^2}{1-\kappa\xi\phi^2}
      \nonumber\\
      &
      -
      \left(
         \left(\sec x + \sin x\frac{\kappa\xi\phi\Phi}{1-\kappa\xi\phi^2}\right)^2
         -
         \sin^2x\frac{\kappa^2\xi^2\phi^2\Pi^2}{\left(1-\kappa\xi\phi^2\right)^2}
      \right)
      \hat{A}
   \Bigg)
   .
\end{eqnarray}
The Misner-Sharp mass is a quasi-local mass, which we use to compute
the mass enclosed by an apparent horizon (when/if it forms) and
the total mass of the spacetime\footnote{We evaluate 
(\ref{eq:misner_sharp_mass}) at $x=\pi/2-\Delta x$, where $\Delta x$
is the size of our grid spacing as we cannot evaluate directly at $x=\pi/2$
due to the factor of $(\cos x)^{-1}$ in that expression.}.
\section{Results\label{sec:results}}

In this section we consider the evolution of a compact scalar field pulse, 
defined at $t = 0$ by the initial data
\begin{equation} 
   \label{eq:gaussian_ID}
   \phi(0,x) 
   = 
   0
   , 
   ~~~ 
   \Phi(0,x) 
   = 
   0
   , 
   ~~~ 
   \Pi(0,x) 
   = 
   \epsilon \exp\Big( - \frac{4 \tan^2 x}{\pi^2 \sigma^2} \Big)
   ,
\end{equation}
with $\sigma = \frac{1}{16}$ \cite{Bizon:2011gg}. 
Bizon and Rostworowski \cite{Bizon:2011gg} 
conjectured that AdS spacetime with a minimally coupled scalar
field was unstable to
forming black holes with this class of initial data for all $\epsilon$.
They numerically found that the
amplitude $\epsilon$ set the timescale to black hole formation:
for large enough $\epsilon$, a black hole would form during the
first implosion of the scalar field; smaller $\epsilon$ values would
lead to the scalar field bouncing off the origin and outer boundary
until a black hole would form during implosion.
The process by which field profiles with
arbitrarily small $\epsilon$ are able to eventually collapse is referred to as
``weak turbulence'', referencing the turbulent transfer of energy to small
spatial scales commonly studied in fluid dynamics.  In this section,
we first present results on weak-field initial data, and follow the transfer of
energy to increasingly shorter time/spatial scales.
We then consider the dynamics of strong-field initial data, 
which collapse to a black hole after few bounces.  
In each case, 
we investigate the impact the nonminimal coupling, $\xi$, has on the solutions.

\begin{figure}[h]
\centering
\includegraphics[height=0.5\textheight]{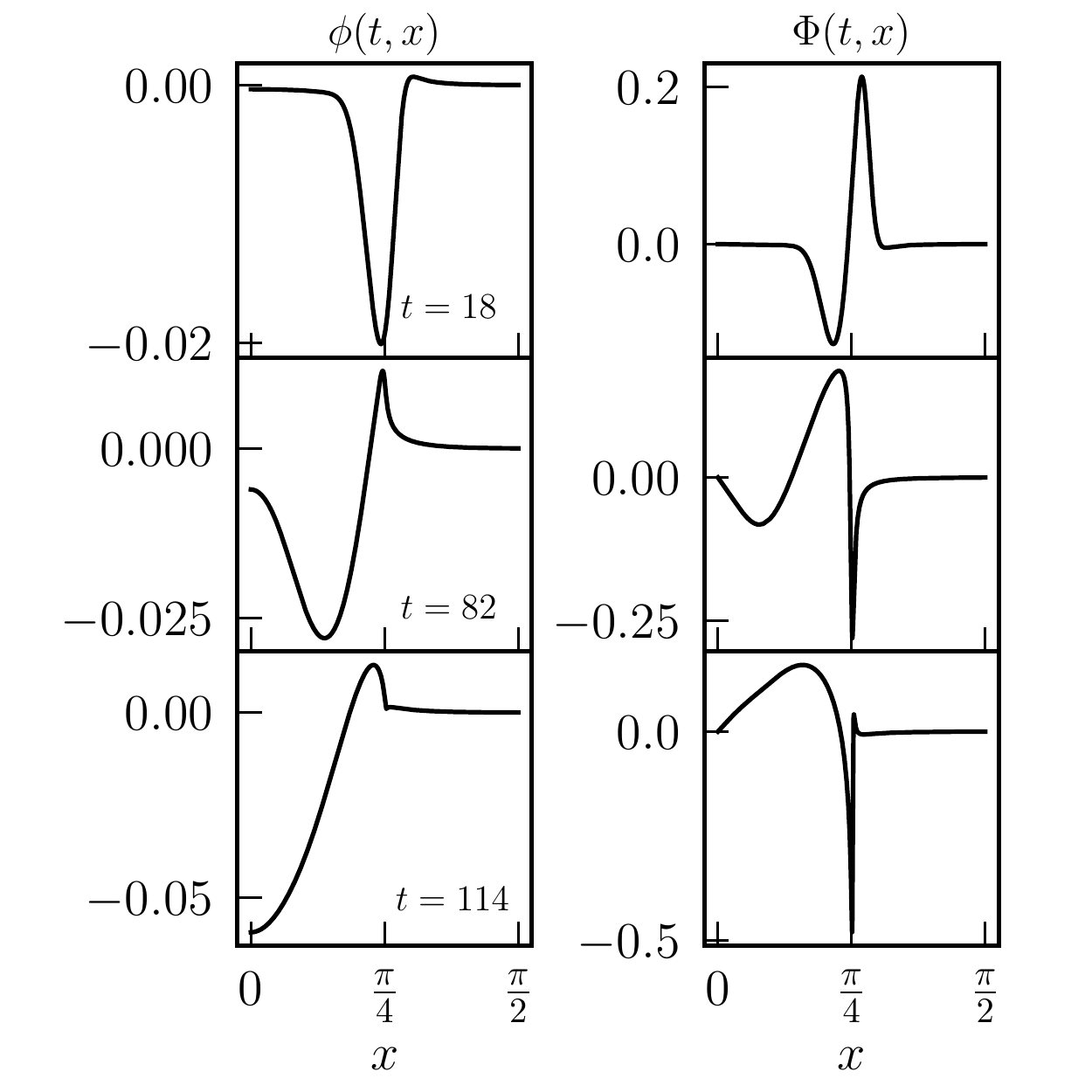}
\caption{Plot of the scalar field $\phi$ and its spatial derivative 
   $\Phi$ from a simulation with initial data (\ref{eq:gaussian_ID}), 
   $\epsilon = 6 \sqrt{2}$ and $\xi = \frac{1}{6}$.  
   Each panel is a snapshot of $\phi$ and $\Phi$ at times when the 
   pulse is moving toward the origin but is roughly centered about the 
   middle of the domain $x = \frac{\pi}{4}$; starting from the top, 
   each lower panel occurs successively later in the evolution.  
   As time progresses, one can see that energy is transferred from the 
   left side of the compact $\phi$ profile toward the right edge, 
   forming a steep jump centered near $x = \frac{\pi}{4}$.  
   The sharpening of this jump corresponds to growth of the spatial derivative 
   $\Phi$, which grows by a factor of five from $t = 18$ to $t = 114$.} 
   \label{fig:phi_vs_t}
\end{figure}

\begin{figure}[h]
\centering
\includegraphics[width=\textwidth]{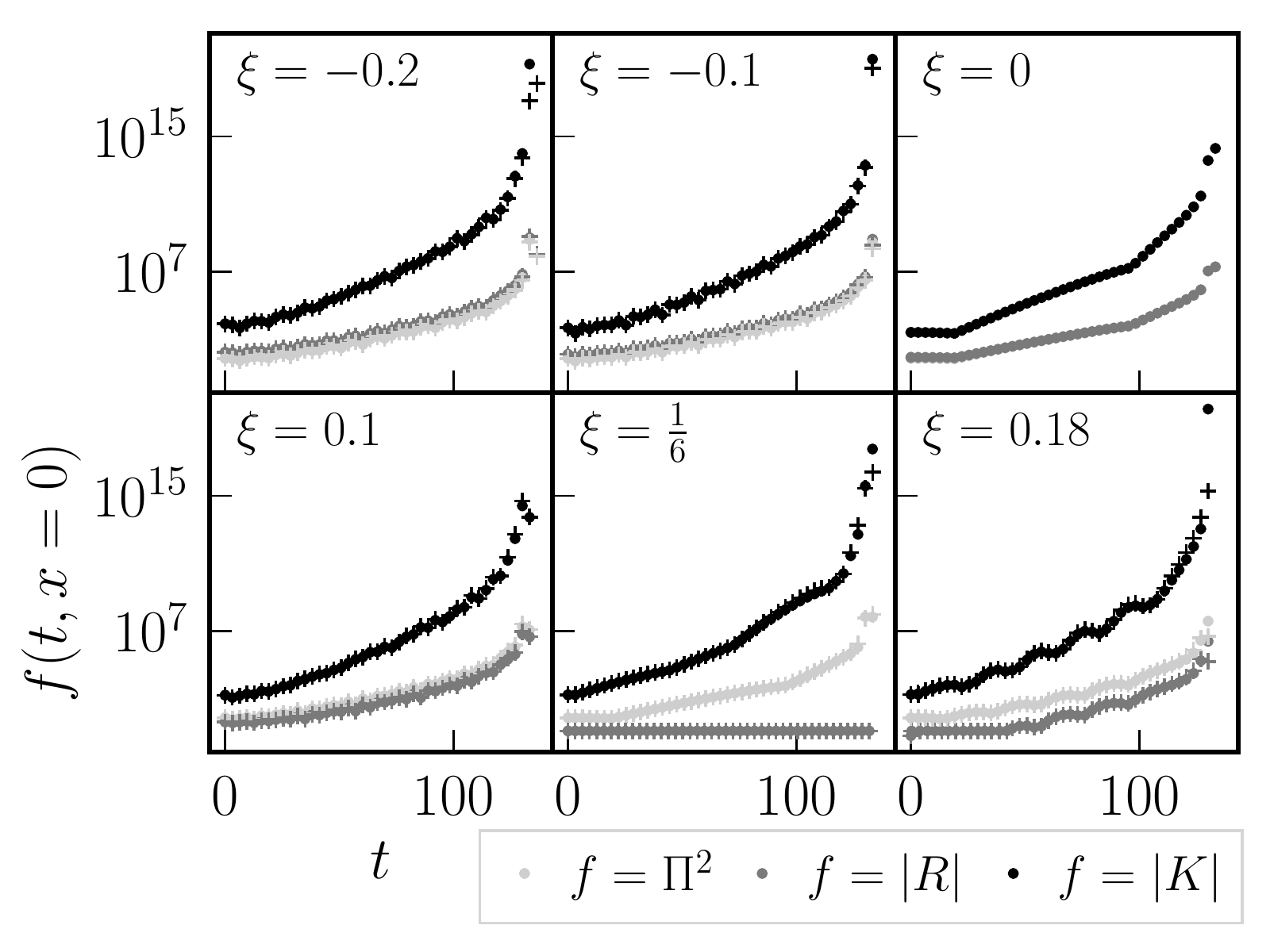}
\caption{Comparison of three Jordan-frame diagnostics---$\Pi^2, |R|, 
   |K|$ evaluated at the origin---for different values of $\xi$ 
   for simulations starting from initial data (\ref{eq:gaussian_ID}) 
   with $\epsilon = 6 \sqrt{2}$, shown in dots.  
   For the sake of comparison, the same diagnostics are overlaid in with the 
   same colors except with cross-shaped markers for simulations with a minimal 
   coupling $\xi = 0$ but equivalent mass as given by (\ref{eq:meff_definition}), 
   e.g. in the top-left panel the dots represent $\Pi^2, |R|, |K|$ 
   from a simulation with $\xi = -0.2$, and the crosses are $\Pi^2, |R|, |K|$ 
   from a simulation with $\xi = 0, m^2 = 2.4$.  
   For $\xi = 0$, the dot and cross simulations are identical, 
   so the crosses are omitted.  
   As in \cite{Bizon:2011gg}, $f(t,0)$ above traces out a continuous curve 
   in each case above, but we plot only the peak value achieved in each 
   implosion through the origin to de-clutter the figure.  
   Note that the growth of $f(t,0)$ is always non-monotonic for 
   $\xi = -0.2, -0.1, 0.1, 0.18$, but is monotonic after an early nearly 
   growth-free phase for the two special cases $\xi = 0, \xi = \frac{1}{6}$.  
   Note also that for $\xi = \frac{1}{6}$, $R$ is constant; 
   see (\ref{eq:Jordan_R}).} \label{fig:envelope_plot}
\end{figure}

\begin{figure}[h]
\centering
\includegraphics[width=0.75\textwidth]{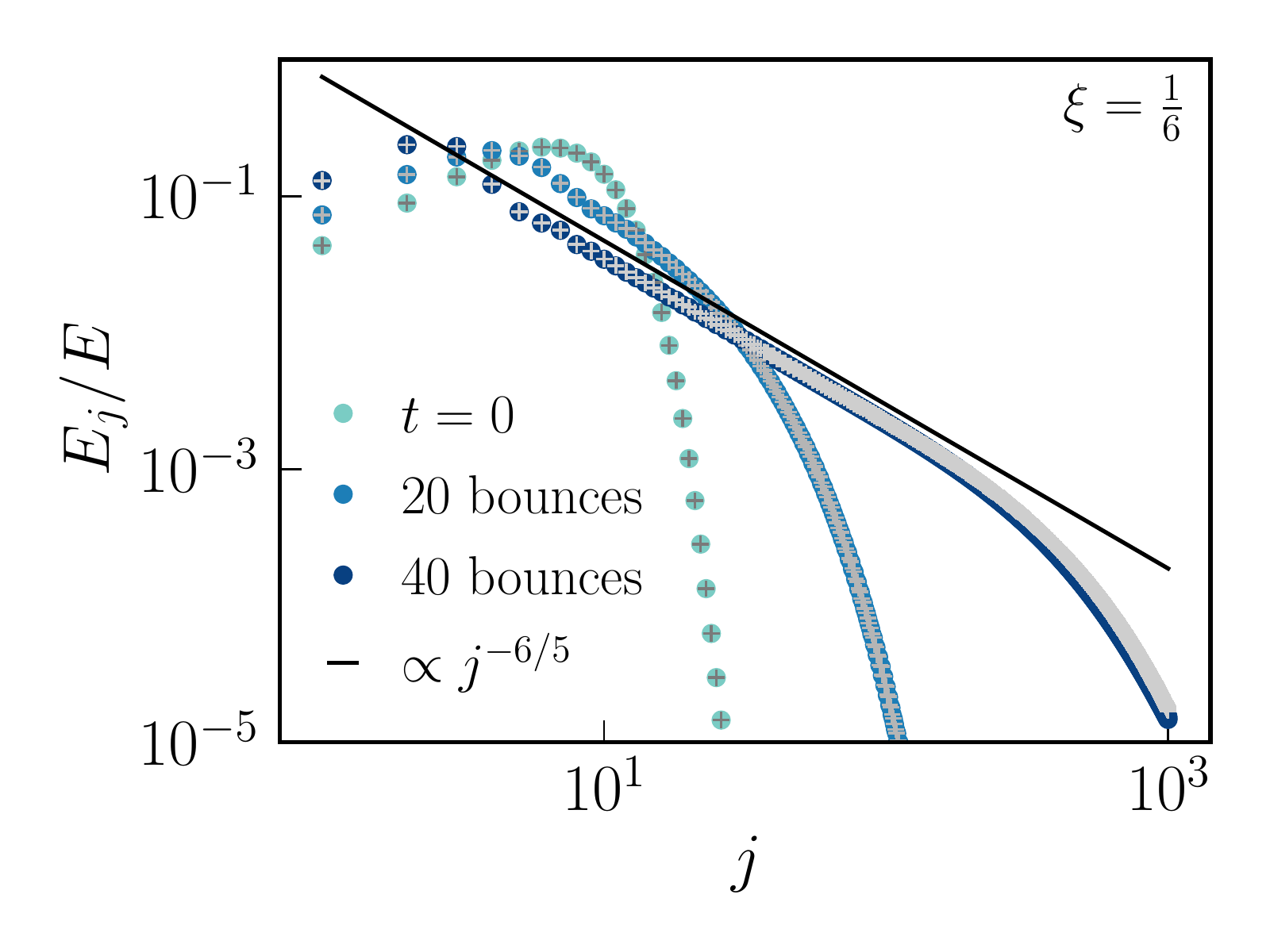}
\caption{Energy per linearized mode $E_j$ (\ref{eq:mode_solutions}) as a 
   function of $j$ for a simulation with $\xi = \frac{1}{6}$, 
   $\epsilon = 6 \sqrt{2}$ in blue dots.  
   Also included are points from a simulation with $\xi = 0$ 
   and the analogous mass given by (\ref{eq:meff_definition}), namely $m^2 = -2$, 
   in gray plus signs.  
   The solution shows that at early times, only low-$j$ modes are 
   populated and higher $j$ modes are exponentially suppressed.  
   As time passes, energy is transferred to the higher $j$ modes, 
   and at late times (a couple bounces before horizon formation) 
   a set of these higher $j$ modes follow a power law 
   $\propto j^{-6/5}$ before exponentially decaying at very high $j$
   (see \cite{Buchel:2012uh}).
   Here $\xi = \frac{1}{6}$ is chosen as a representative case; 
   for $\xi = -0.2, -0.1, 0, 0.1, 0.18$ the figure is essentially 
   identical.
   } \label{fig:mode_energy}
\end{figure}

Beginning with weak-field initial data, all of our simulations are consistent
with the notion that black holes can form for initial data 
(\ref{eq:gaussian_ID}) with arbitrarily small amplitudes $\epsilon$.
In Figs. \ref{fig:phi_vs_t}-\ref{fig:mode_energy} we focus on simulations
starting from (\ref{eq:gaussian_ID}) with 
$\epsilon = 6 \sqrt{2}$, which bounces off of the AdS boundary 42 
times before forming a horizon. 
We find evidence of the transfer of energy to shorter scales long
before the formation of the final black hole. 
In Fig. \ref{fig:phi_vs_t},
we show the scalar field profile from a simulation with $\xi = \frac{1}{6}$.  
One can see that the $\phi$ profile, which is characterized at early times by a negative
lobe and a positive one, slowly transfers energy from the former to the latter.
In particular, the outer edge of the positive lobe forms a very sharp jump,
which indicates the that energy is concentrated near that point.  
The formation of a
sharp feature is picked up by the spatial gradient of the scalar field, $\Phi$,
which exhibits secular growth near the jump in $\phi$ at $x \sim \frac{\pi}{4}$.

The growth of gradients is shown for various values of $\xi$ in Fig. 
\ref{fig:envelope_plot} from simulations starting from initial data 
(\ref{eq:gaussian_ID}) with $\epsilon = 6 \sqrt{2}$.  Three different
metrics are shown: the growth of $\Pi^2$, roughly showing the enhancement
of $\dot{\phi}$ with each successive bounce, as well as the magnitudes
of the Jordan-frame Ricci and Kretschmann scalars, all evaluated at the
origin of spherical symmetry.  As is discussed in \cite{Bizon:2011gg},
the continuous profile (for example, $\Pi^{2}(t,0)$) takes a
complicated form at late times, with many rapid oscillations, 
so we only show a single point
corresponding to the maximum value of each quantity achieved in a 
cycle of the scalar field imploding through the origin.  The figure shows
that all of the aforementioned quantities grow at roughly the same rate 
as a function of the coupling constant, though it is clear that the extent 
to which this growth is \textit{monotonic} is affected by the choice of 
$\xi$.  In particular, it seems that the cases with $\xi = 0, \frac{1}{6}$
grow roughly monotonically at late times, whereas $\xi = -0.2, -0.1, 0.1$
all develop nontrivial patterns of local maxima and minima.  The case
$\xi = 0.18$, somewhat near the BF bound $\xi = 0.1875$, (\ref{eq:BF_bound}),
shows a pronounced oscillation pattern superimposed over the secular growth 
of $\Pi^{2}, R, K$.  Overlaid on each panel in plus-shaped markers are the
same results except for simulations with the equivalent mass 
(\ref{eq:meff_definition}) and a minimal coupling $\xi = 0$.  The massive,
minimally-coupled results are effectively identical to those with the
non-minimal coupling until late times (within a few bounces of collapse),
indicating that the aforementioned growth of gradients is indeed occurring
in the weak-field regime where the nonminimal coupling acts like a mass
term in the scalar wave equation (\ref{eq:meff_definition},
\ref{eq:decoupled_KGE}).

Fig. \ref{fig:mode_energy} shows the energy (\ref{eq:mode_energy}) of the
solution projected onto the orthonormal basis of solutions to the linearized
wave equation on a fixed $AdS_4$ background (\ref{eq:mode_solutions}).  It
turns out that the rate of energy transfer is effectively identical for
various values of $\xi \in [-0.2, 0.18]$, so only a single case 
($\xi = \frac{1}{6}$) is shown.  Again, overlaid in plus-shaped markers are
the same results except for a minimally-coupled massive scalar field with
mass given by (\ref{eq:meff_definition}) corresponding to $\xi = \frac{1}{6}$.
At early times, the projection of the solution onto modes with high indices $j$ 
(corresponding to $e_j$ with rapid oscillations in $x$) are exponentially 
suppressed. As time goes on, these modes get populated and the exponential
decay is replaced by a power-law with slope $\approx -6/5$, in agreement with
that found in \cite{Maliborski:2013via}.  The population of high-$j$ modes
is cut off by the formation of an apparent horizon after 42
bounces, only shortly later than the latest time (darkest blue dots) shown 
in Fig. \ref{fig:mode_energy}.

\begin{figure}[h]
\centering
\includegraphics[width=1.0\textwidth]{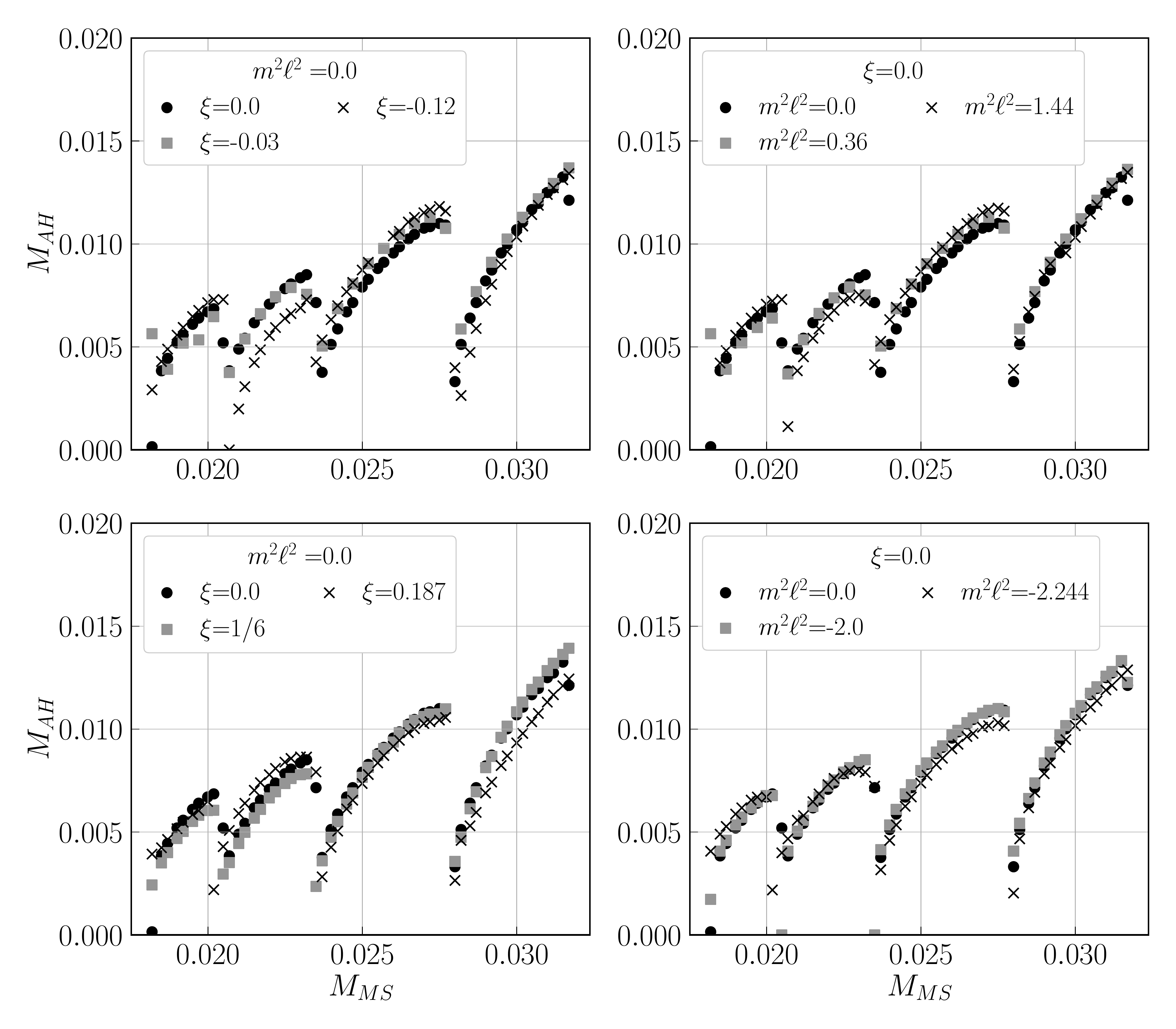}
   \caption{Mass of the initial apparent horizon $M_{AH}$ versus the total
   mass of the spacetime $M_{MS}$ (see Sec.~\ref{sec:ah_and_ms})
   for Gaussian initial data (\ref{eq:gaussian_ID}), where all
   masses are evaluated in the Jordan frame.
   The left panels show how the black hole mass changes for a massless nonminimally
   coupled scalar field.
   The right panels show how the black hole mass changes for a minimally coupled massive
   scalar field, where the masses are chosen to agree with the effective masses 
   (\ref{eq:meff_definition}) of the cases in the left panels.
   The finger-like structures correspond to (from right to left) solutions which
   bounce zero, one, two, and three times off of the AdS boundary before forming
   an apparent horizon.
   } 
   \label{fig:mass_comparison}
\end{figure}

We next consder strong-field initial data, which collapse to a black hole after only
a few bounces off the AdS boundary, and focus on the formation of the apparent horizon.  
In Fig.~\ref{fig:mass_comparison} we compare 
the mass enclosed in the initial apparent horizon (AH), $M_{AH}$, that forms and the 
total Misner-Sharp mass of the spacetime $M_{MS}$ (see Sec.~\ref{sec:ah_and_ms}) for 
Gaussian initial data for different values of $\xi$ and for a massive 
minimally coupled scalar field with mass $m^2=m_{eff}^2$.
In this figure, it is clear that the AH mass is not monotonic with the total
mass of the spacetime; instead, one sees a set of four finger-like clusters
which are monotonic in $M_{MS}$, with sudden jumps in $M_{AH}$ in between them.  Each finger
corresponds to (from right to left) cases which collapse after zero, one, two, 
and three bounces of the scalar field off of the AdS boundary.
We find that the initial data (\ref{eq:gaussian_ID}) eventually forms apparent
horizons in a qualitatively similar way for nonminimally coupled scalar fields
as is seen for minimally coupled fields
\cite{Bizon:2011gg}.
The masses of the initial apparent horizons are not dramatically affected by
the nonminimal coupling nor by a nonzero mass term with $\xi = 0$ when compared
against those for a minimally coupled, massless field.

\section{The scalar field-viscous fluid analogy
\label{sec:scalar_fluid_analogy}
}

There is a formal analogy between 
the stress-energy tensor of a minimally coupled scalar field and that 
of a perfect fluid \cite{1980ApJ...235.1038M,landau2013fluid}.
This analogy can be made rigorous for 
irrotational barotropic fluids, where the relativistic Euler equations 
can be rewritten exactly as a scalar wave equation for a suitably defined 
scalar field \cite{Fajman:2020tcf,Rodnianski:2009de}, 
but often in other contexts a scalar field 
is used for the matter model and the results are 
\textit{interpreted} by analogy to a fluid
(see, e.g. \cite{Aditya:2019wql,Trodden:2004st,Liddle2000} in the
context of inflationary cosmology).
We adopt the latter approach here, except with the modification that 
the nonminimal coupling term in the action (\ref{eq:Jordan_action}) 
results in the presence of terms in the stress-energy tensor 
(\ref{eq:Jordan_Tab}) which map to \textit{viscous corrections} 
to the relativistic fluid \cite{Madsen:1988ph,Faraoni:2012hn}.

The fact that the nonminimal coupling parameter $\xi$ introduces terms 
which act as an effective viscosity in the fluid analog is of 
particular interest when considering the ``weakly turbulent'' 
instability of AdS spacetimes, where black holes are formed from weak
initial data through the transfer of energy to small scales.  
If the scalar field is indeed acting as an effective viscous fluid, 
one might expect the effective viscosity introduced by $\xi$ 
to combat the transfer of energy to small scales, perhaps slowing 
or even halting the formation of a black hole.  
In this section we first briefly review the scalar field-fluid analogy,
evaluate the extent to which the scalar field actually behaves as a fluid
in our numerical solutions, 
and show that the scalar's effective viscosity does not noticeably 
alter the qualitative features of the solutions.

To derive the scalar field-fluid analogy,
we first decompose the stress-energy tensor 
(assumed only to be a symmetric two-tensor) with respect to a unit 
timelike four-vector $u^{a}$, and then
write it as \cite{Eckart1940,Kovtun:2012rj}
\begin{equation} \label{eq:Tab_decomp}
   T^{ab} 
   = 
   \mathcal{E} u^{a} u^{b} 
   + 
   \mathcal{P} \Delta^{ab} 
   + 
   \mathcal{Q}^{a} u^{b} 
   + 
   \mathcal{Q}^{b} u^{a} 
   + 
   \mathcal{T}^{ab}
   ,
\end{equation}
where
\begin{eqnarray} \label{eq:Tab_decomp_terms}
   &\mathcal{E} 
   \equiv
   u_{a} u_{b} T^{a b}, 
   ~~~ 
   \mathcal{P} \equiv
   \frac{1}{3} \Delta_{ab} T^{ab}, 
   ~~~ 
   \mathcal{Q}^{a} 
   \equiv
   -\Delta^{a b} u^{c} T_{b c} 
   \nonumber \\
   &
   \mathcal{T}^{ab} 
   \equiv
   T^{<ab>}
   \equiv 
   \frac{1}{2} \Big( \Delta^{a c} \Delta^{b d} + \Delta^{a d} \Delta^{b c} - \frac{2}{3} \Delta^{a b} \Delta^{c d} \Big) T_{c d},
   \\
   &
   \Delta^{ab} \equiv g^{ab} + u^a u^b
   \nonumber
   .
\end{eqnarray}
As written, substituting (\ref{eq:Tab_decomp_terms}) into 
(\ref{eq:Tab_decomp}) simply yields the identity $T^{ab} = T^{ab}$.  
Fluid models are defined by replacing (\ref{eq:Tab_decomp_terms}) 
with a set of constitutive relations defining the components 
$\mathcal{E}, \mathcal{P}, \mathcal{Q}^{a}, \mathcal{T}^{ab}$ 
in terms of a set of hydrodynamic variables derived from equilibrium
thermodynamics, and the four-vector $u^{a}$ is defined to be the
(equilibrium) flow velocity of the fluid.

Rather than defining constitutive relations, we instead notice
that for a fluid model, $\mathcal{E}$ plays the role of the energy
density, $\mathcal{P}$ is the pressure, $\mathcal{Q}^{a}$ is the
heat flux vector, and $\mathcal{T}^{ab}$ incorporates the effects
of shear viscosity.  Though in principle all of the aforementioned
terms can acquire non-equilibrium (dissipative) corrections,
$\mathcal{Q}^{a}$ and $\mathcal{T}^{ab}$ model 
purely \emph{non-equilibrium} effects, and thus vanish for a fluid
in equilibrium (see e.g. \cite{Kovtun:2012rj}).
Constructing the ``fluid analog'' of a given non-fluid stress-energy tensor 
then consists of taking the projections (\ref{eq:Tab_decomp_terms}) 
and interpreting them as the energy density, pressure, heat flux,
and shear viscosity of a relativistic fluid.

In order to perform the decomposition 
(\ref{eq:Tab_decomp}-\ref{eq:Tab_decomp_terms}) 
for the scalar-tensor theory (\ref{eq:Jordan_Tab}-\ref{eq:Jordan_Sab}), 
one must define a timelike unit four-vector 
to identify with the flow velocity of the fluid.  
The usual definition \cite{Madsen:1988ph}, 
\begin{equation} \label{eq:scalar_u_a}
   u^{a} 
   \equiv 
   \frac{\nabla^{a} \phi}{\sqrt{-\nabla_{c} \phi \nabla^{c} \phi}} 
   \equiv 
   \frac{\nabla^{a} \phi}{N}
   ,
\end{equation}
is well-defined so long as the gradient of the
scalar field remains timelike.
One can now perform the decomposition by taking the 
projections (\ref{eq:Tab_decomp_terms}), which yields 
\begin{eqnarray}
   \mathcal{E}_{s} &= \frac{1}{1 - \kappa \xi \phi^2} \Big( \frac{1}{2} N^2 + \frac{\Lambda}{\kappa} - 2 \xi N \phi \nabla_b u^b \Big) \label{eq:En} \\
   \mathcal{P}_{s} &= \frac{1}{1 - \kappa \xi \phi^2} \Big( \frac{1}{2} N^2 - \frac{\Lambda}{\kappa} + 2 \xi \Big[ \phi u^c \nabla_{c} N + \frac{2}{3} \phi N \nabla_{c} u^c - N^2 \Big] \Big) \label{eq:P} \\
   \mathcal{Q}^{a}_{s} &= \frac{2 \xi N \phi}{1 - \kappa  \xi \phi^2} \, u^c \nabla_c u^a \label{eq:Qa} \\
   \mathcal{T}^{ab}_{s} &= - 2 \frac{\xi  N \phi}{1 - \kappa \xi \phi^2} \, \sigma^{ab}, \label{eq:STT_Tab}
\end{eqnarray}
where $\sigma^{ab} = \nabla^{<a} u^{b>}$ (see (\ref{eq:Tab_decomp_terms})) and
each component is given a subscript $s$ to denote 
that it is derived from the scalar field stress-energy tensor 
(\ref{eq:Jordan_Tab}-\ref{eq:Jordan_Sab}).
We see that the nonminimal coupling adds corrections to 
$\mathcal{E}, \mathcal{P}$, but most significantly makes the effective
heat flow vector and shear viscosity nonzero.
In this sense, the effects of the nonminimal coupling $\xi \neq 0$ 
map onto non-equilibrium (viscous) effects in the analog fluid.  

\begin{figure}
\centering
\includegraphics[width=0.6\columnwidth]{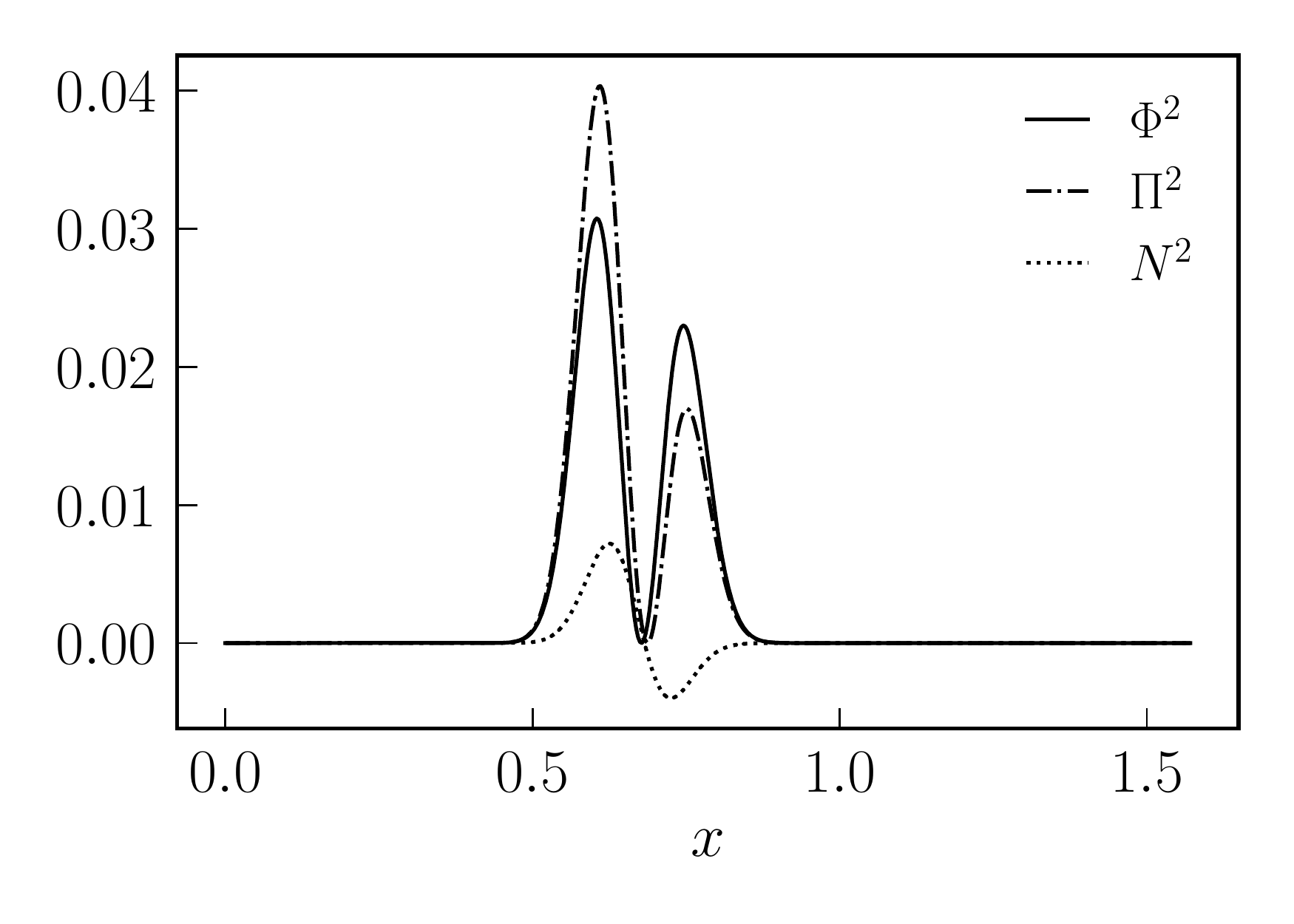}
\caption{Snapshot from an evolution starting from Gaussian initial data 
   (\ref{eq:gaussian_ID}) with $\epsilon = 10$, at a time when the field 
   is concentrated near the center of the domain.  
   Note that $\Phi^2 > \Pi^2$ over the entire region $x \gsim 0.75$, 
   implying the four-vector identified with the flow velocity in the 
   fluid analogy (\ref{eq:scalar_u_a}) is spacelike rather than timelike.
   } \label{fig:fluid_norm}
\end{figure}

Ultimately, we find that the viscous fluid analogy is only of limited use 
in understanding our results.
Computing $N^2$ in the coordinate basis used in our simulations, we find
\begin{equation} \label{eq:N_in_basis}
N^2 = \Omega^2 \frac{\cos^2 x}{\ell^2} \hat{A} \, \big( \Pi^2 - \Phi^2 \big),
\end{equation}
which is negative if $\Phi^2 > \Pi^2$.  
The effective four-velocity is almost never timelike across the entire spatial 
domain due to the fact that $\Pi$ decays faster than $\Phi$ 
approaching the AdS boundary 
($\Phi \sim \rho^{\Delta-1}, \Pi \sim \rho^{\Delta}$, see (\ref{eq:bc_fields})). 
Since $\Delta\geq3/2>0$, we expect that 
\emph{the fluid interpretation 
      can only be valid across 
      the entire spatial domain when $\Phi$ is compactly supported
      away from the AdS boundary.}
Fig. \ref{fig:fluid_norm} shows $\Phi^2, \Pi^2$, and $N^2$ 
for an evolution starting from Gaussian initial data (\ref{eq:gaussian_ID}) 
with $\epsilon = 10$ at a point in time when the field is concentrated in 
the center of the domain---note that $N^2$ is negative from $x \sim 0.75$ 
all the way to the outer boundary $x = \frac{\pi}{2}$.

\begin{figure}
\centering
\includegraphics[width=\columnwidth]{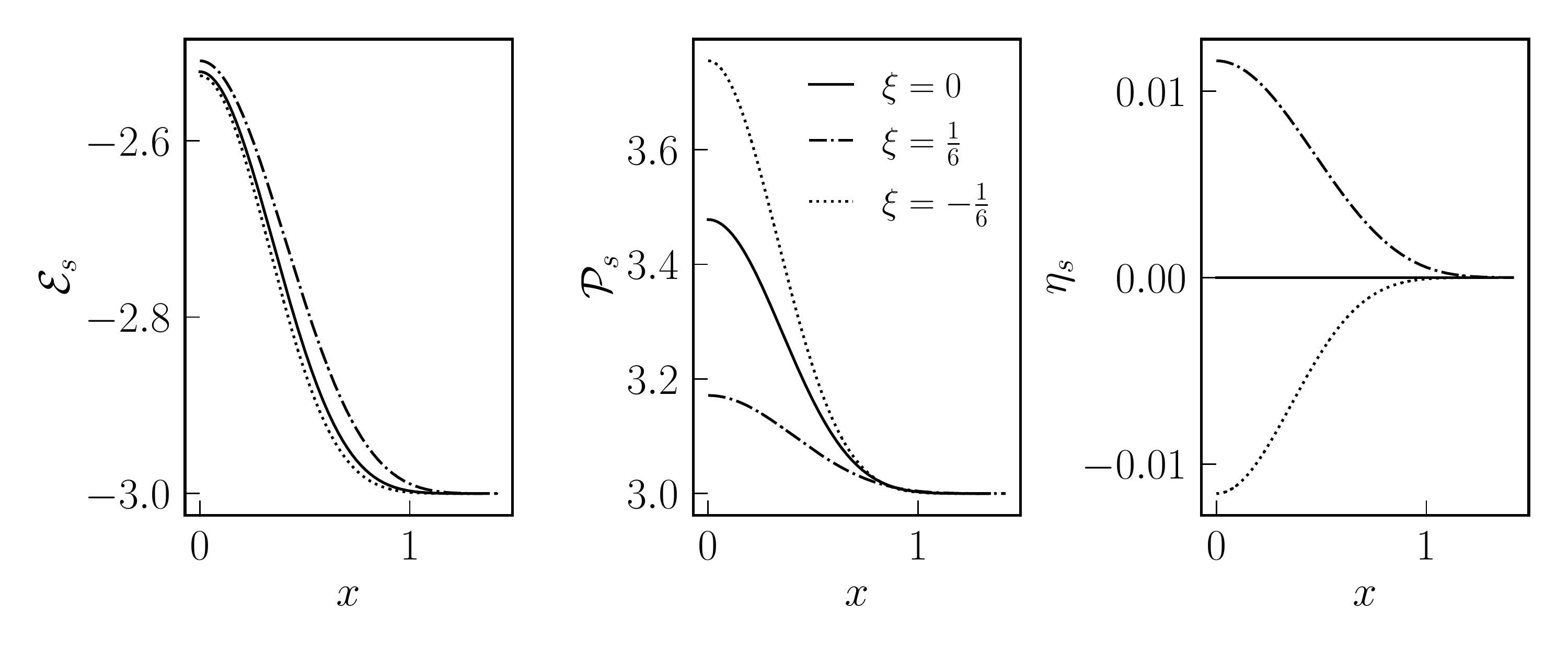}
\caption{Components of the fluid analog (\ref{eq:En}-\ref{eq:STT_Tab}) 
   from an evolution starting from mode-0 initial data (\ref{eq:mode_solutions}) 
   of the form (\ref{eq:mode0_ID}) with $\epsilon = 1$, 
   at a time when the four-vector $u^a$ (\ref{eq:scalar_u_a}) 
   is timelike almost everywhere.  
   Shown in the three panels are the total energy density seen by 
   a co-moving observer $\mathcal{E}_s$ (\ref{eq:En}), 
   the pressure $\mathcal{P}_s$ (\ref{eq:P}), 
   and the effective shear viscosity $\eta_s$ (\ref{eq:eta_s}), 
   for scalar fields with $\xi = 0, \frac{1}{6}, -\frac{1}{6}$.  
   Note that the energy density is negative, 
   which is unphysical and amounts to a violation of the weak energy condition.  
   The pressure is positive, as expected, 
   and seems to be enhanced for $\xi < 0$ and diminished for $\xi > 0$.  
   The effective shear viscosity changes sign during the evolution, 
   and the sign at a given time is determined by the sign of the product 
   $\xi \phi$ (\ref{eq:eta_s}).
   } 
   \label{fig:fluid_comps}
\end{figure}

That said, in the weak-field limit ($\phi \ll 1$) the mode solutions 
(\ref{eq:mode_solutions}) are time-periodic and will, at times, 
have $\Phi = 0 \, \forall \, x$.  
We show the components $\mathcal{E}_{s}, \mathcal{P}_{s}$ (\ref{eq:En}-\ref{eq:P})
as well as the effective shear viscosity 
(defined by comparison of $\mathcal{T}^{ab} = - 2 \eta \sigma^{ab}$ 
to (\ref{eq:STT_Tab}) for the viscous fluid \cite{Kovtun:2012rj})
\begin{equation} \label{eq:eta_s}
\eta_s \equiv \frac{\xi  N \phi}{1 - \kappa \xi \phi^2},
\end{equation}
of the fluid analog from an evolution with initial data based on the 
$n = 0$ mode of (\ref{eq:mode_solutions}), namely
\begin{equation} \label{eq:mode0_ID}
\phi(0, x) = 0, ~~~ \Phi(0, x) = 0, ~~~ \Pi(0, x) = \epsilon \cos(x)^{\Delta}
\end{equation}
with $\epsilon = 1$ at a time when $u^a$ (\ref{eq:scalar_u_a}) 
is timelike everywhere in Fig \ref{fig:fluid_comps}.  
Even at times when the fluid analog is well-defined
(i.e. where $u^a$ is timelike),
we find that the total energy density is negative ($\mathcal{E}_{s} < 0$)---which
signals a violation of the weak energy condition for an observer 
co-moving with  the ``fluid''---and the effective shear viscosity changes 
sign with the  scalar field, periodically attaining unphysical negative values.

In summary, the scalar field-viscous fluid analogy is not of much utility
for the case considered here, as the quantity typically identified with
the effective fluid's four-velocity (\ref{eq:scalar_u_a}) does not remain
timelike in the presence of significant spatial anisotropy.  
For the case shown here when the analog \emph{is} well-defined, 
the effective fluid displays 
a negative energy density and the shear viscosity changes sign in time.  
Similar effects were seen in \cite{Faraoni:2021lfc}, though they find reasonable behavior 
when defining certain thermodynamic quantities (such as the temperature).  
We find that the effective viscosity (\ref{eq:eta_s}) 
does not inhibit (or enhance) the weakly
turbulent instability of $AdS_4$ spacetime in our simulations.

\section{Discussion\label{sec:discussion}}

In this study we have numerically investigated the instability of
asymptotically $AdS_4$ spacetimes for a nonminimally coupled scalar
field model. We find that the qualitative behavior
of our solutions do not differ from those of minimally coupled
massive and massless scalar fields:
from small ``generic'' initial scalar field perturbations, energy is
transferred over time to increasingly smaller scales, 
until a black hole is formed.
While we work in the Einstein frame of the theory, where the nonminimal
coupling is absorbed into the metric via a Weyl rescaling, 
we measure the formation of black holes and the transfer of energy 
in the Jordan frame, where the scalar field couples directly
to the Ricci scalar.

Our work does not address the stability of black holes in the
theory we consider.
In the Jordan frame, the theory in principle allows for violations of
the NCC, which underlies the classical black hole area theorem 
\cite{hawking_area_original}.
It would be interesting to consider the stability of 
asymptotically AdS spacetime in horizon-penetrating
coordinates, as one would then be able to evolve the
dynamics of the solution beyond apparent horizon formation
to determine if it is eventually enveloped by an event horizon.
Critical collapse solutions have been studied for nonminimally
coupled scalar fields in asymptotically flat spacetimes \cite{Choptuik_1992_nmc},
although not to our knowledge for asymptotically AdS spacetimes;
it would be interesting to determine the effect (if any)
of the cosmological constant on such critical collapse solutions.
We leave to future work a more thorough study of the dynamics of
collapse for values of $\xi$ very close to the Breitenlohner-Freedman bound 
($\xi = \frac{3}{16} = 0.1875$).
While we were able to perform numerical evolutions of $\xi$ somewhat near 
the bound (up to $\xi = 0.187$), 
it would be interesting to determine the behavior of solutions for theories
which saturate that bound.

In Sec. \ref{sec:scalar_fluid_analogy} we considered the interpretation 
of the scalar field's stress-energy tensor as that of a 
relativistic viscous fluid.  
This analogy is of particular interest here, 
as one could conjecture that the viscosity in the fluid analog 
(which is introduced by the nonminimal coupling $\xi$) 
would hamper the transfer of energy to small scales 
characterizing the AdS instability.  
Unfortunately the analogy breaks down in the case of interest, 
as it is only applicable when the gradient of the scalar field is timelike, 
which generically is not the case in parts of the domain due 
to spatial anisotropy in the field configuration.  
That said, the question of whether or not true physical viscosity inhibits 
the instability of AdS is an open question, 
and it may be addressed using a legitimate relativistic 
viscous fluid theory such as, for example, 
M\"uller-Israel-Stewart theory \cite{Muller_1967,Israel:1976tn,Israel_1979} 
or Bemfica-Disconzi-Noronha-Kovtun theory \cite{Bemfica:2020zjp,Kovtun:2019hdm}.
In such a study, the presence of viscosity should combat the gradient-sharpening effect of 
the AdS instability, and it would be quite interesting to determine which effect wins out 
as a function of the ``amount'' of viscosity in the model.  
In principle, there are three possibilities: 
(1) an infinitesimal amount of viscosity makes AdS spacetime
stable to small enough perturbations
(it adds a ``mass gap'' to the black holes that form from perturbative
initial data); 
(2) a finite amount of viscosity is required for stability, 
in which case the threshold would be interesting to investigate; or 
(3) viscosity has no impact on or only slows the growth of gradients 
and horizons still form.  
The answer to this question would further elucidate the structure of 
the instability, 
and would also serve to clarify the relationship between AdS 
``turbulence'' and that experienced in fluid flows.

\ack

   We thank Frans Pretorius and Elias Most for helpful conversations
   about the relativistic fluid analogy for scalar fields. 
   During the course of this work JLR was 
   supported by STFC Research Grant No. ST/V005669/1.
   Some of the simulations presented in this article were performed 
   on computational 
   resources managed and supported by Princeton Research Computing, 
   a consortium of groups including the Princeton Institute for 
   Computational Science and Engineering (PICSciE) and the 
   Office of Information Technology's High Performance Computing Center and 
   Visualization Laboratory at Princeton University.
   This work also made use of the Cambridge Service for Data Driven
   Discovery (CSD3), 
   part of which is operated by the University of Cambridge Research
   Computing on behalf of the STFC DiRAC HPC Facility (www.dirac.ac.uk)

\newpage
\appendix
\section{Numerical methods and convergence tests \label{sec:numerical_methods}}

We solve the evolution equations
(\ref{eq:Einstein_Pi_eqn}), (\ref{eq:Pi_defn}),
the constraint equations
(\ref{eq:Einstein_Aprime_eqn}) and (\ref{eq:Einstein_delta_eqn}),
and the spatial derivative of (\ref{eq:Pi_defn}):
\begin{equation}
   \dot{\Phi}
   =
   \left(\hat{A}e^{-\hat{\delta}}\Pi\right)'
   ,
\end{equation}
using finite difference methods.  We evolve $\phi,\Phi,\Pi$ in time using the 
method of lines with the standard fourth-order Runge-Kutta method, and use 
fourth-order finite difference stencils to approximate spatial derivatives.

As is described in \cite{Bizon:2011gg,Maliborski:2013via}, the main challenge in 
solving these equations lies in maintaining stability at the origin and 
at the AdS boundary. 
To help maintain stability, following \cite{Buchel:2012uh} we 
rescale the scalar field and its derivatives via
\begin{eqnarray} \label{eq:rescaled_matter_fields}
   \tilde{\phi}
   \equiv&
   \frac{1}{\cos^{\Delta-1} x}\phi
   ,\\
   \tilde{\Pi}
   \equiv&
   \frac{1}{\cos^{\Delta-1} x}\Pi
   ,\\
   \tilde{\Phi}
   \equiv&
   \frac{1}{\cos^{\Delta-2} x}\Phi
   ,
\end{eqnarray}
and these variables are evolved rather than $\phi, \Phi, \Pi$.  
This is done
because the fields $\phi, \Phi, \Pi$ decay at rates inversely proportional to 
$\xi$ approaching the AdS boundary, implying larger field values near 
$x = \frac{\pi}{2}$ for large $\xi$ than for small $\xi$; this, in turn, results
in a loss of stability for $\xi > 0$.  The rescaled fields always have the
falloff  $\tilde{\phi},\tilde{\Pi},\tilde{\Phi} \sim \rho$, resulting in significantly
improved stability for $\xi > 0$.  
The definitions (\ref{eq:rescaled_matter_fields}), when combined
with the falloff behavior (\ref{eq:origin_falloff}), (\ref{eq:bc_fields})
implies the boundary conditions
\begin{eqnarray}
&\phi'(0) = \Phi(0) = \Pi'(0) = 0 \\
&\phi \Big(\frac{\pi}{2} \Big) = \Phi \Big(\frac{\pi}{2} \Big) = \Pi \Big(\frac{\pi}{2} \Big) = 0.
\end{eqnarray}

Rather than using forward- or backward-biased finite difference stencils near the origin
and AdS boundary respectively, we use a fourth-order centered stencil and set the field values 
lying outside of the domain using the parity of the field at the boundary: 
namely, $\phi, \Pi$ are even and $\Phi$ is odd at the origin, and 
$\phi, \Phi, \Pi$ are all odd at the outer boundary.  
This means that if a stencil requires fields at a point $-x_i < 0$, for example, 
then $\phi(-x_i) = \phi(x_i)$, $\Pi(-x_i) = \Pi(x_i)$, and 
$\Phi(-x_i) = - \Phi(x_i)$; for points $x_i > \frac{\pi}{2}$, the odd parity of 
the fields $f \in \{\phi, \Phi, \Pi\}$ implies $f(\frac{\pi}{2}-x_i) = -f(\frac{\pi}{2}+x_i)$.

It turns out that, even with fourth-order spatial derivative stencils,
the truncation error in the $\Pi$ evolution equation (\ref{eq:Einstein_Pi_eqn}) is 
large enough near the origin and the AdS boundary to destabilize the 
numerical  scheme. Thus, instead of solving (\ref{eq:Einstein_Pi_eqn}) at the first
and last interior gridpoints---i.e. the gridpoint adjacent to the 
gridpoint at the origin and the gridpoint adjacent to the
gridpoint at the AdS boundary---we 
use cubic spline interpolation to set the value of $\Pi$.  
A ``natural cubic spline'' gives an interpolated value of a function $f$ at position 
$x$ using values of $f$ and its derivatives at so-called knot points 
$x_{i}, x_{j}$ (where $x_i \leq x \leq x_j$) \cite{atkinson2004elementary}:
\begin{eqnarray} \label{eq:cubic_spline}
\fl
f(x) 
= 
\frac{
	(x_j-x)^3 f''(x_i) + (x - x_i)^3 f''(x_j)
}{
	6 (x_j - x_i)
} 
+ 
\frac{
	(x_j - x) f(x_i) + (x - x_i) f(x_j)
}{
	x_j - x_i
}
\nonumber\\
-
\frac{1}{6}\left(x_j-x_i\right)\left[
\left(x_j-x\right)f''(x_i)
+
\left(x-x_i\right)f''(x_j)
\right]
,
\end{eqnarray}
where in this case we take $f = \Pi$, $x = x_1$, $x_i = x_0$, and $x_j = x_2$.
The second derivative terms $\Pi''$ are computed using fourth-order centered
finite difference stencils, making use of the parity of $\Pi$ to set field 
values when the stencil extends past the boundary of the domain.

Spline interpolation (\ref{eq:cubic_spline}) turns out to be especially effective
at damping spurious numerical oscillations, and as a result is essential to the
stability of solutions which have many bounces off of the boundaries (such as
the 42-bounce data shown in Fig. \ref{fig:envelope_plot}, and the many-bounce 
evolutions shown in \cite{Bizon:2011gg,Maliborski:2013via}).  For these 
simulations, the standard fourth-order Runge-Kutta method is used to integrate 
the constraint equations (\ref{eq:Einstein_Aprime_eqn}) and 
(\ref{eq:Einstein_delta_eqn}).  This method requires values of the fields in 
between spatial gridpoints (at locations $x_{k + \frac{1}{2}}$), so we use 
natural cubic spline interpolation (\ref{eq:cubic_spline}) with 
$x = x_{k+\frac{1}{2}}$, $x_{i} = x_{k}$, and $x_{j} = x_{k+1}$.  For the study
of apparent horizon formation (Fig. \ref{fig:mass_comparison}), the simulations 
of interest form apparent horizons within only a couple of bounces, so a simpler 
second-order Runge-Kutta method is used to integrate the constraint equations; 
this method does not require field values at half-gridpoints, so cubic spline
interpolation is only used for $\Pi$ at the gridpoints near the inner and outer
boundaries.
This being said, we were unable to stably numerically
evolve simulations for values of
$\xi$ that saturate the BF bound, $\xi=3/16$.

To solve the constraints, we integrate (\ref{eq:Einstein_delta_eqn})
for $\hat{\delta}$ inward from the AdS boundary to the origin.
We solve for $\hat{A}$ by first defining the Einstein frame
Misner-Sharp mass function \cite{Buchel:2012uh}
\begin{equation}
   \label{eq:einstein_frame_mass}
   1
   -
   \frac{2\hat{M}_{MS}(t,x)}{\hat{R}_{area}}
   -
   \frac{1}{3}\hat{R}_{area}^2\Lambda
   =
   \hat{g}^{ab}\hat{\nabla}_a\hat{R}_{area}\hat{\nabla}_b\hat{R}_{area}
   ,
\end{equation}
where $\hat{R}_{area}=\ell \tan x$.
Substituting for $\hat{R}$, from Eq.~(\ref{eq:einstein_frame_mass}) we find
that 
\begin{equation}
   \label{eq:hatA_as_hatM}
   \hat{A}\left(t,x\right)
   =
   1
   -
   2\frac{\cos^3x}{\sin x}\frac{\hat{M}_{MS}\left(t,x\right)}{\ell}
   ,
\end{equation}
where our definition for the Einstein frame Misner-Sharp mass differs
from that of Buchel et. al. \cite{Buchel:2012uh} by a factor of $2/\ell$.
Using (\ref{eq:hatA_as_hatM}),
the constraint (\ref{eq:Einstein_Aprime_eqn}) then becomes 
\begin{equation}
   \fl
   \hat{M}_{MS}'
   =
   \frac{\kappa \ell}{2} \frac{\sin^2x}{\cos^{2}x}
   \left(
      \frac{1}{2}\left[
         \Pi^2
         +
         \Phi^2
      \right]
   \left[1-2\frac{\cos^3}{\sin x}\frac{\hat{M}_{MS}}{\ell}\right] 
      q\left(\phi\right)
      -
      \frac{3\xi}{\cos^2x}
      \;z\left(\phi\right)
   \right)
   .
\end{equation}
We integrate this expression rather than (\ref{eq:Einstein_Aprime_eqn}),
as doing so results in significantly improved numerical stability; 
we then apply (\ref{eq:einstein_frame_mass}) to recover $\hat{A}$.

\begin{figure}[h]
\centering
\includegraphics[width=0.6\columnwidth]{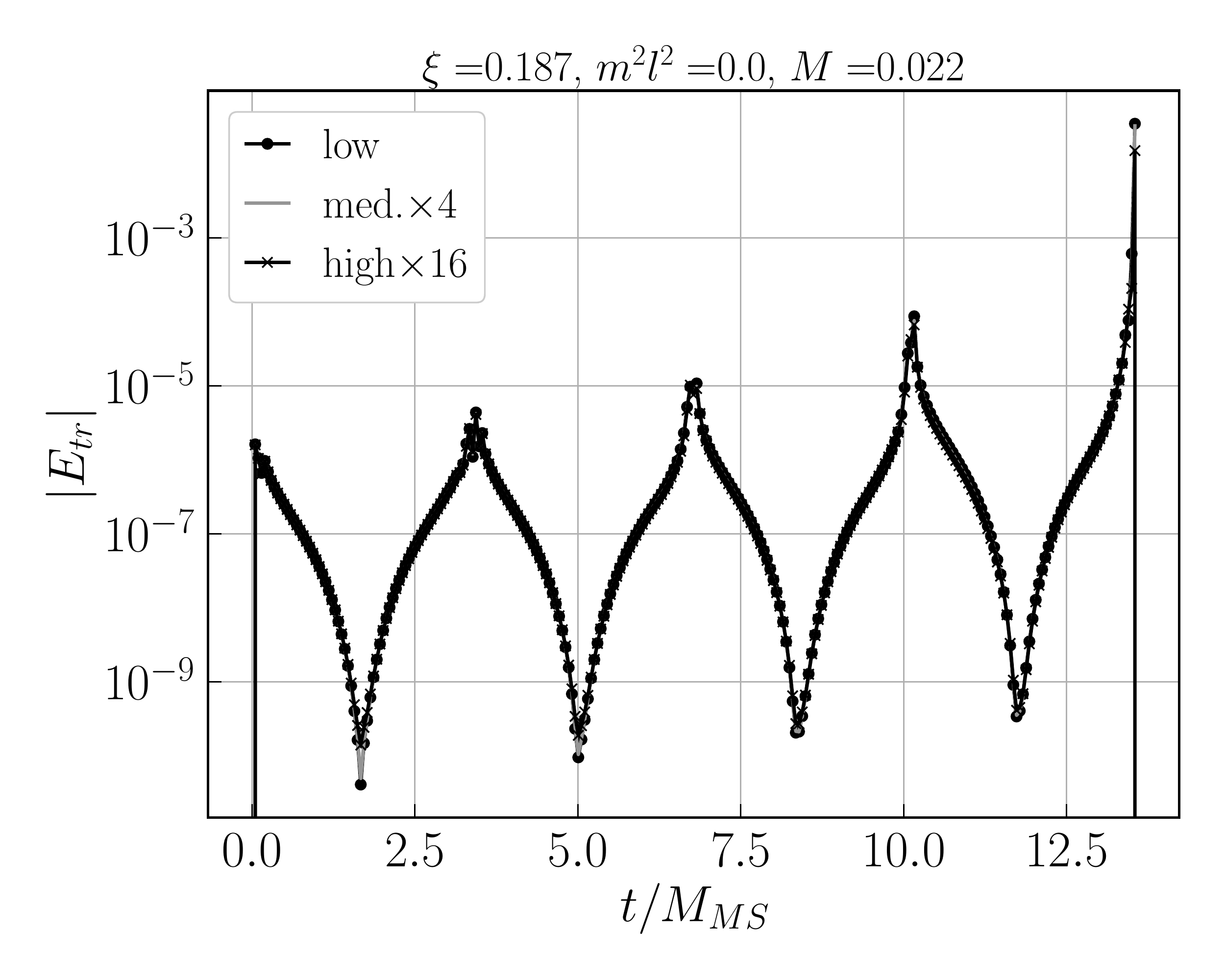}
\caption{Plot of the one norm of the independent residual
   (\ref{eq:Einstein_Adot_eqn}), which we denote by the ``$tr$'' component
   of the Einstein equations, $E_{tr}$. The ``low'' resolution corresponds to
   $2^{12}$ radial grid points, and the ``med.'' and ``high'' resolutions have
   twice and four times that base grid resolution. We rescale the one-norm
   by $4$ and $16$ in the plot, which shows second-order convergence, which
   is consistent with the stencils we used in our code to compute
   the constraint equations.
   Here $M_{MS}$ is the total Misner-Sharp mass of the spacetime.
   } 
   \label{fig:convergence_indep_res}
\end{figure}

\begin{figure}[h]
\centering
\includegraphics[width=0.6\columnwidth]{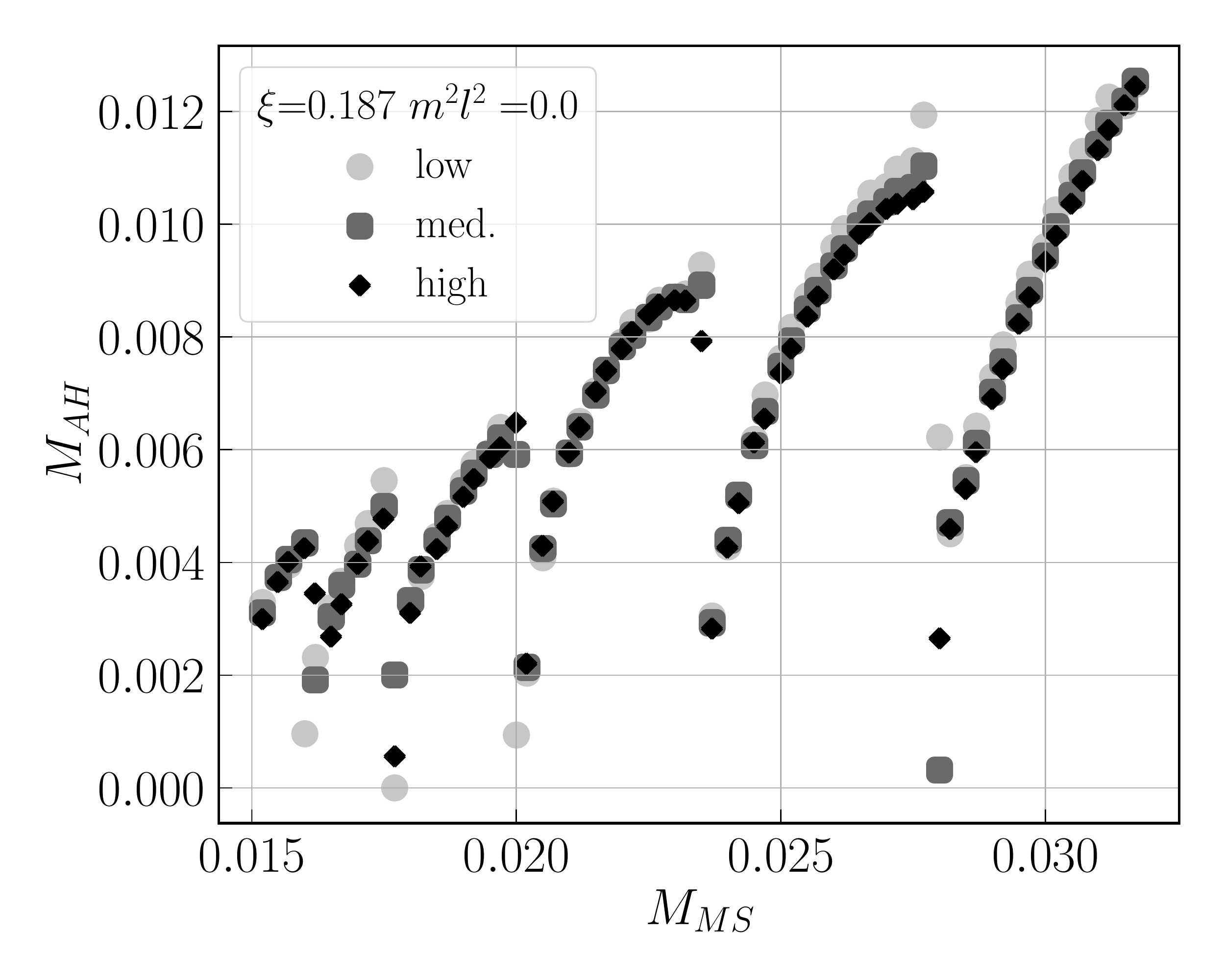}
\caption{Misner-Sharp mass of the first apparent horizon versus the total
   Misner-Sharp mass. We see that we are able to resolve the formation
   of apparent horizons for the range of initial data we considered.
   The ``low'' resolution run has $2^{12}$ radial grid points, while
   the ``med.'' and ``high'' runs have twice and four times that resolution.
   The scatter in the grid points near the boundaries between the finger-like clusters indicates
   the difficulty in resolving the masses of very small black holes.
   The data presented in Fig.~\ref{fig:mass_comparison} 
   were run at the ``high'' resolution.
   } 
   \label{fig:convergence_mass_ah}
\end{figure}

To check the convergence of our code, we evaluated the equation
for $\dot{\hat{A}}$, Eq.~(\ref{eq:Einstein_Adot_eqn}), 
which is unused by our solution algorithm,
and verified that it goes to zero with higher resolution at a rate consistent
with the accuracy of the numerical scheme used.
We plot a representative convergence test in Fig.~\ref{fig:convergence_indep_res}
for a run that had three bounces, $\xi=0.187$, and total Misner-Sharp mass
of $0.022$. We find that the independent residual goes to zero at the expected
rate for a scheme which is second-order in the grid spacing, in accordance with
the fact that the scheme used solves the constraint equations 
(\ref{eq:Einstein_Aprime_eqn}-\ref{eq:Einstein_delta_eqn}) using a second-order
Runge-Kutta integrator.  In Fig. \ref{fig:convergence_mass_ah} we show that the 
apparent horizon mass does not vary significantly with numerical resolution for 
evolutions with $\xi=0.187$ (the least numerically stable case we consider), 
implying that the dynamics are sufficiently converged at the resolutions used.


\section{Jordan-frame Kretschmann scalar} \label{sec:Kretschmann}
As a diagnostic we compute the Kretschmann scalar 
$K \equiv R_{a b c d} R^{a b c d}$ derived from the Jordan frame metric 
(\ref{eq:Jordan_metric}).  
We only use $K$ at the origin of the spherically symmetric spacetime, 
which is
\begin{eqnarray}
\fl
   K(t,0) = \frac{1}{l^4 \left(1 - \kappa \xi \phi^2\right)^2} \Bigg( 4 \kappa ^3 \xi ^3 \phi ^6 \Big(-\hat{A}''^2-e^{2 \hat{\delta} } \kappa  \xi  \left(\dot{\phi}^2 (\hat{A}''-2 \hat{\delta}''+10)+2 \dot{\hat{\delta}} \dot{\phi} \phi''+2 \ddot{\phi} \phi''\right) \nonumber\\
\fl
   +4 \hat{A}'' \hat{\delta}''-4 \hat{A}''+3 e^{4 \hat{\delta} } \kappa  \xi  (\dot{\hat{\delta}} \dot{\phi}+\ddot{\phi})^2-4 \hat{\delta}''^2+8 \hat{\delta}''+3 \kappa  \xi  \phi''^2-20\Big) \nonumber\\
\fl
   +2 \kappa ^2 \xi ^2 \phi ^4 \Big(3 \hat{A}''^2+2 \hat{A}'' \left(e^{2 \hat{\delta} } \kappa  \xi  \dot{\phi}^2-6 \hat{\delta}''+6\right) \nonumber\\
\fl
   +4 \kappa  \xi  \left(2 e^{2 \hat{\delta} } \phi'' (\dot{\hat{\delta}} \dot{\phi}+\ddot{\phi})-3 e^{4 \hat{\delta} } (\dot{\hat{\delta}} \dot{\phi}+\ddot{\phi})^2+e^{2 \hat{\delta} } \dot{\phi}^2 \left(3 e^{2 \hat{\delta} } \kappa  \xi  \dot{\phi}^2-\hat{\delta}''+7\right)-3 \phi''^2\right)+12 ((\hat{\delta}''-2) \hat{\delta}''+5) \Big) \nonumber\\
\fl
   +4 \kappa  \xi  \phi ^2 \Big(-\hat{A}''^2+\hat{A}'' \left(e^{2 \hat{\delta} } \kappa  \xi  \dot{\phi}^2+4 \hat{\delta}''-4\right) \nonumber\\
\fl
   +\kappa  \xi  \left(-2 e^{2 \hat{\delta} } \phi'' (\dot{\hat{\delta}} \dot{\phi}+\ddot{\phi})+3 e^{4 \hat{\delta} } (\dot{\hat{\delta}} \dot{\phi}+\ddot{\phi})^2+2 e^{2 \hat{\delta} } \dot{\phi}^2 \left(3 e^{2 \hat{\delta} } \kappa  \xi  \dot{\phi}^2-\hat{\delta}''+1\right)+3 \phi''^2\right)-4 ((\hat{\delta}''-2) \hat{\delta}''+5) \Big) \nonumber\\
\fl
   +\kappa ^4 \xi ^4 \phi ^8 \left(\hat{A}''^2-4 (\hat{A}''+2) \hat{\delta}''+4 \hat{A}''+4 \hat{\delta}''^2+20\right) 
   \nonumber\\
   \fl
   -12 \kappa ^3 \xi ^3 \phi ^5 \left(e^{2 \hat{\delta} } (\dot{\hat{\delta}} \dot{\phi}+\ddot{\phi})-\phi''\right) \left(\hat{A}''+2 e^{2 \hat{\delta} } \kappa  \xi  \dot{\phi}^2-2 \hat{\delta}''+6\right) \nonumber\\
\fl
   +4 \kappa ^2 \xi ^2 \phi ^3 \left(e^{2 \hat{\delta} } \left(3 (\hat{A}''-2 \hat{\delta}''+6) (\dot{\hat{\delta}} \dot{\phi}+\ddot{\phi})-4 \kappa  \xi  \dot{\phi}^2 \phi''\right)-3 \phi'' (\hat{A}''-2 \hat{\delta}''+6)\right) \nonumber\\
\fl
   -4 \kappa ^4 \xi ^4 \phi ^7 (\hat{A}''-2 \hat{\delta}''+6) \left(\phi''-e^{2 \hat{\delta} } (\dot{\hat{\delta}} \dot{\phi}+\ddot{\phi})\right) \nonumber\\
\fl
   +4 \kappa  \xi  \phi  \left(e^{2 \hat{\delta} } (\dot{\hat{\delta}} \dot{\phi}+\ddot{\phi}) \left(-\hat{A}''+6 e^{2 \hat{\delta} } \kappa  \xi  \dot{\phi}^2+2 \hat{\delta}''-6\right)+\phi'' \left(\hat{A}''-2 \left(e^{2 \hat{\delta} } \kappa  \xi  \dot{\phi}^2+\hat{\delta}''-3\right)\right)\right) \nonumber\\
\fl
   -4 e^{2 \hat{\delta} } \kappa  \xi  \dot{\phi}^2 (\hat{A}''-2 \hat{\delta}''+6)+(\hat{A}''-2 \hat{\delta}'')^2+4 (\hat{A}''-2 \hat{\delta}''+5)+12 e^{4 \hat{\delta} } \kappa ^2 \xi ^2 \dot{\phi}^4 \Bigg).
\end{eqnarray}

\newpage
\bibliographystyle{iopart-num}
\bibliography{thebib}

\end{document}